\documentclass[twocolumn]{./aastex6}
\RequirePackage{fix-cm}
\usepackage{epsfig}
\usepackage{hyperref}
\usepackage{color}
\usepackage{xcolor}
\usepackage{fancyvrb}
\usepackage{subfigure}
\usepackage{soul}
\begin{document}

\title{$C^{3}$, A Command-line Catalog Cross-match Tool for Large Astrophysical Catalogs}
\author{Giuseppe Riccio\dag, Massimo Brescia, Stefano Cavuoti, Amata Mercurio}
\affil{INAF Astronomical Observatory of Capodimonte - via Moiariello 16, I-80131 Napoli, Italy}
\email[\dag]{ giuseppe.riccio08@gmail.com}
\author{Anna Maria di Giorgio, Sergio Molinari}
\affil{INAF Istituto di Astrofisica e Planetologia Spaziali - Via Fosso del Cavaliere 100, I-00133 Roma, Italy}


\date{Received 2016 September 06; accepted: 2016 November 14; published 2016 MM DD}

\graphicspath{{./}}

\begin{abstract}
Modern Astrophysics is based on multi-wavelength data organized into large and heterogeneous catalogs. Hence, the need for efficient, reliable and scalable catalog cross-matching methods plays a crucial role in the era of the petabyte scale. Furthermore, multi-band data have often very different angular resolution, requiring the highest generality of cross-matching features, mainly in terms of region shape and resolution. In this work we present $C^{3}$ (Command-line Catalog Cross-match), a multi-platform application designed to efficiently cross-match massive catalogs. It is based on a multi-core parallel processing paradigm and conceived to be executed as a stand-alone command-line process or integrated within any generic data reduction/analysis pipeline, providing the maximum flexibility to the end-user, in terms of portability, parameter configuration, catalog formats, angular resolution, region shapes, coordinate units and cross-matching types. Using real data, extracted from public surveys, we discuss the cross-matching capabilities and computing time efficiency also through a direct comparison with some publicly available tools, chosen among the most used within the community, and representative of different interface paradigms. We verified that the $C^{3}$ tool has excellent capabilities to perform an efficient and reliable cross-matching between large data sets. Although the elliptical cross-match and the parametric handling of angular orientation and offset are known concepts in the astrophysical context, their availability in the presented command-line tool makes $C^{3}$ competitive in the context of public astronomical tools.
\end{abstract}

\keywords{methods: data analysis  -- catalogs -- techniques: miscellaneous  -- surveys}

\maketitle

\section{Introduction}

In the last decade we entered the data-intensive era of astrophysics, where the size of data has rapidly increased, reaching in many cases dimensions overcoming the human possibility to handle them in an efficient and comprehensible way. In a very close future petabytes of data will be the standard and, to deal with such amount of information, also the data analysis techniques and facilities must quickly evolve. For example the current exploration of petabyte-scale, multi-disciplinary astronomy and Earth observation synergy, by taking the advantage from their similarities in data analytics, has issued the urgency to find and develop common strategies able to achieve solutions in the data mining algorithms, computer technologies, large scale distributed database management systems as well as parallel processing frameworks \citep{aspera2012}.

Astrophysics is one of the most involved research fields facing with this data explosion, where the data volumes from the ongoing and next generation multi-band and multi-epoch surveys are expected to be so huge that the ability of the astronomers to analyze, cross-correlate and extract knowledge from such data will represent a challenge for scientists and computer engineers. To quote just a few, the ESA Euclid space mission will acquire and process about 100 GBday$^-1$ over at least 6 years, collecting a minimum amount of about $200$TB of data \citep{laureijs2014}; Pan-STARRS \citep{kaiser2004} is expected to produce more than $100$TB of data; the GAIA space mission will build a $3D$ map of the Milky Way galaxy, by collecting about one petabyte of data in five years \citep{douglas2007}; the Large Synoptic Survey Telescope (\citealt{ivezic2009}) will provide about $20$TB/night of imaging data for ten years and petabytes/year of radio data products. Many other planned instruments and already operative surveys will reach a huge scale during their operational lifetime, such as KiDS (Kilo-Degree Survey; \citealt{deJong+15_KIDS_paperI}), DES (Dark Energy Survey, \citealt{annis2013}), Herschel-ATLAS \citep{valiante2015,varga2016}, Hi-GAL \citep{molinari2016}, SKA \citep{braun2015} and E-ELT \citep{martins2014}.

The growth and heterogeneity of data availability induce challenges on cross-correlation algorithms and methods. Most of the interesting research fields are in fact based on the capability and efficiency to cross-correlate information among different surveys. This poses the consequent problem of transferring large volumes of data from/to data centers, \textit{de facto} making almost inoperable any cross-reference analysis, unless to change the perspective, by moving software to the data \citep{cavuoti2012}.

Furthermore, observed data coming from different surveys, even if referred to a same sky region, are often archived and reduced by different systems and technologies. This implies that the resulting catalogs, containing billions of sources, may have very different formats, naming schemas, data structures and resolution, making the data analysis to be a not trivial challenge. Some past attempts have been explored to propose standard solutions to introduce the uniformity of astronomical data quantities description, such as in the case of the Uniform Content Descriptors of the Virtual Observatory \citep{ivoa2005}.

One of the most common techniques used in astrophysics and fundamental prerequisite for combining multi-band data, particularly sensible to the growing of the data sets dimensions, is the cross-match among heterogeneous catalogs, which consists in identifying and comparing sources belonging to different observations, performed at different wavelengths or under different conditions. This makes cross-matching one of the core steps of any standard modern pipeline of data reduction/analysis and one of the central components of the Virtual Observatory \citep{malkov2012}.

The massive multi-band and multi-epoch information, foreseen to be available from the on-going and future surveys, will require efficient techniques and software solutions to be directly integrated into the reduction pipelines, making possible to cross-correlate in real time a large variety of parameters for billions of sky objects. Important astrophysical questions, such as the evolution of star forming regions, the galaxy formation, the distribution of dark matter and the nature of dark energy, could be addressed by monitoring and correlating fluxes at different wavelengths, morphological and structural parameters at different epochs, as well as by opportunely determining their cosmological distances and by identifying and classifying peculiar objects. In such context, an efficient, reliable and flexible cross-matching mechanism plays a crucial role.
In this work we present $C^{3}$ (\textit{Command-line Catalog Cross-match\footnote{The $C^{3}$ tool and the user guide are available at the page \url{http://dame.dsf.unina.it/c3.html}.}}, \citealt{riccio2016}), a tool to perform efficient catalog cross-matching, based on the multi-thread paradigm, which can be easily integrated into an automatic data analysis pipeline and scientifically validated on some real case examples taken from public astronomical data archives. Furthermore, one of major features of this tool is the possibility to choose shape, orientation and size of the cross-matching area, respectively, between elliptical and rectangular, clockwise and counterclockwise, fixed and parametric. This makes the $C^{3}$ tool easily tailored on the specific user needs.

The paper is structured as follows: after a preliminary introduction, in Sec.~\ref{SEC:techniques} we perform a summary of main available techniques; in Sec.~\ref{sect:c3design}, the design and architecture of the $C^{3}$ tool is described; in sections \ref{sect:config} and \ref{sect:optimization}, the procedure to correctly use $C^{3}$ is illustrated with particular reference to the optimization of its parameters; some tests performed in order to evaluate $C^{3}$ performance are shown in Sec.~\ref{sect:performances}; finally, conclusions and future improvements are drawn in Sec.~\ref{sect:conclusion}.

\section{Cross-matching techniques}\label{SEC:techniques}

Cross-match can be used to find detections surrounding a given source or to perform one-to-one matches in order to combine physical properties or to study the temporal evolution of a set of sources.

The primary criterion for cross-matching is the approximate coincidence of celestial coordinates (positional cross-match). There are also other kinds of approach, which make use of the positional mechanism supplemented by statistical analysis used to select best candidates, like the bayesian statistics \citep{budavari2008}. In the positional cross-match, the only attributes under consideration are the spatial information. This kind of match is of fundamental importance in astronomy, due to the fact that the same object may have different coordinates in various catalogs, for several reasons: measurement errors, instrument sensitivities, calibration, physical constraints, etc.

In principle, at the base of any kind of catalog cross-match, each source of a first catalog should be compared with all counterparts contained in a second catalog. This procedure, if performed in the naive way, is extremely time consuming, due to the huge amount of sources. Therefore different solutions to this problem have been proposed, taking advantage of the progress in computer science in the field of multi-processing and high performing techniques of sky partitioning. Two different strategies to implement cross-matching tools basically exist: web and stand-alone applications.

Web applications, like OpenSkyQuery \citep{nieto2006}, or CDS-Xmatch \citep{pineau2011}, offer a portal to the astronomers, allowing to cross-match large astronomical data sets, either mirrored from worldwide distributed data centers or directly uploadable from the user local machine, through an intuitive user interface. The end-user has not the need to know how the data are treated, delegating all the computational choices to the backend software, in particular for what is concerning the data handling for the concurrent parallelization mechanism. Other web applications, like ARCHES \citep{motch2015}, provide dedicated script languages which, on one hand, allow to perform complex cross-correlations while controlling the full process but, on the other hand, make experiment settings quite hard for an astronomer.
Basically, main limitation of a web-based approach is the impossibility to directly use the cross-matching tool in an automatic pipeline of data reduction/analysis. In other words, with such a tool the user cannot design and implement a complete automatic procedure to deal with data. Moreover, the management of concurrent jobs and the number of simultaneous users can limit the scalability of the tool. For example, a registered user of CDS-Xmatch has only $500$MB disk space available to store his own data (reduced to $100$MB for unregistered users) and all jobs are aborted if the computation time exceeds 100 minutes \citep{boch2014}. Finally, the choice of parameters and/or functional cases is often limited in order to guarantee a basic use by the end-users through short web forms (for instance, in CDS-Xmatch only equatorial coordinate system is allowed).

Stand-alone applications are generally command-line tools that can be run on the end-user machine as well as on a distributed computing environment. A stand-alone application generally makes use of APIs (Application Programming Interfaces), a set of routines, protocols and tools integrated in the code. There are several examples of available APIs, implementing astronomical facilities, such as STIL\footnote{\url{http://www.star.bris.ac.uk/~mbt/stil/}} \citep{taylor2006}, and astroML\footnote{\url{http://www.astroml.org/}} \citep{vanderplas2012}, that can be integrated by an astronomer within its own source code. However, this requires the astronomer to be aware of strong programming skills. Moreover, when the tools are executed on any local machine, it is evident that such applications may be not able to exploit the power of distributed computing, limiting the performance and requiring the storage of the catalogs on the hosting machine, besides the problem of platform dependency.

On the contrary, a ready-to-use stand-alone tool, already conceived and implemented to embed the use of APIs in the best way, will result an off-the-shelf product that the end-user has only to run. A local command-line tool can be put in a pipeline through easy system calls, thus giving the possibility to the end-user to create a custom data analysis/reduction procedure without writing or modifying any source code. Moreover, being an all-in-one package, i.e including all the required libraries and routines, a stand-alone application can be easily used in a distributed computing environment, by simply uploading the code and the data on the working nodes of the available computing infrastructure.

One of the most used stand-alone tools is STILTS\footnote{\url{http://www.star.bris.ac.uk/~mbt/stilts/}} (STIL Tool Set, \citealt{taylor2006}). It is not only a cross-matching software, but also a set of command-line tools based on the STIL libraries, to process tabular data. It is written in pure Java (almost platform independent) and contains a large number of facilities for table analysis, so being a very powerful instrument for the astronomers. On one hand, the general-purpose nature of STILTS has the drawback to make hard the syntax for the composition of the command line; on the other hand, it does not support the full range of cross-matching options provided by $C^{3}$. In order to provide a more user-friendly tool to the astronomers, it is also available its graphical counterpart, Tool for OPerations on Catalogs And Tables\footnote{\url{http://www.star.bristol.ac.uk/~mbt/topcat/}} (TOPCAT, \citealt{taylor2005}), an interactive graphical viewer and editor for tabular data, based on STIL APIs and implementing the STILTS functionalities, but with all the intrinsic limitations of the graphical tools, very similar to the web applications in terms of use.

Regardless the approach to cross-match the astronomical sources, the main problem is to minimize the computational time exploding with the increasing of the matching catalog size. In principle, the code can be designed according to multi-process and/or multi-thread paradigm, so exploiting the hosting machine features. For instance, \cite{lee2013} evaluated to use a multi-GPU environment, designing and developing their own Xmatch tool, \citep{budavari2013}. Other studies are focused to efficiently cross-match large astronomical catalogs on clusters consisting of heterogeneous processors including both multi-core CPUs and GPUs, (\citealt{jia2015}, \citealt{jia2016}). Furthermore, it is possible to reduce the number of sources to be compared among catalogs, by opportunely partitioning the sky through indexing functions and determining only a specific area to be analyzed for each source. CDS-Xmatch and the tool described in \cite{zhao2009} use Hierarchical Equal Area isoLatitude Pixelisation (HEALPix, \citealt{gorski2005}), to create such sky partition.  \cite{du2014}, instead, proposed a combined method to speed up the cross-match by using HTM (Hierarchical Triangle Mesh, \citealt{kunszt2001}), in combination with HEALPix and by submitting the analysis to a pool of threads.

HEALPix is a genuinely curvilinear partition of the sphere into exactly equal area quadrilaterals of varying shape (see Fig.~3 in \citealt{gorski2005}). The base-resolution comprises twelve pixels in three rings around the poles and equator. Each pixel is partitioned into four smaller quadrilaterals in the next level. The strategy of HTM is the same of HEALPix. The difference between the two spatial-indexing functions is that HTM partitioning is based on triangles, starting with eight triangles, $4$ on the Northern and $4$ on the Southern hemisphere, each one partitioned into four smaller triangles at the next level (see also Fig.~2 in \citealt{du2014}). By using one or both functions combined together, it is possible to reduce the number of comparisons among objects to ones lying in adjacent areas.

Finally OpenSkyQuery uses the \textit{Zones} indexing algorithm to efficiently support spatial queries on the sphere, \citep{gray2006}.

The basic idea behind the \textit{Zones} method is to map the sphere into stripes of a certain height $h$, called zones. Each object with coordinates ($ra$, $dec$) is assigned to a zone by using the formula:

\begin{equation}
 zoneID=dec+90.0/h
\end{equation}

A traditional B-tree index is then used to store objects within a zone, ordered by \textit{zoneID} and right ascension. In this way, the spatial cross-matching can be performed by using bounding boxes (B-tree ranges) dynamically computed, thus reducing the number of comparisons (Fig.~1 in \citealt{nieto2006}). Finally, an additional and expensive test allows to discard false positives.

\begin{figure}
\centering
  \includegraphics[width=6cm]{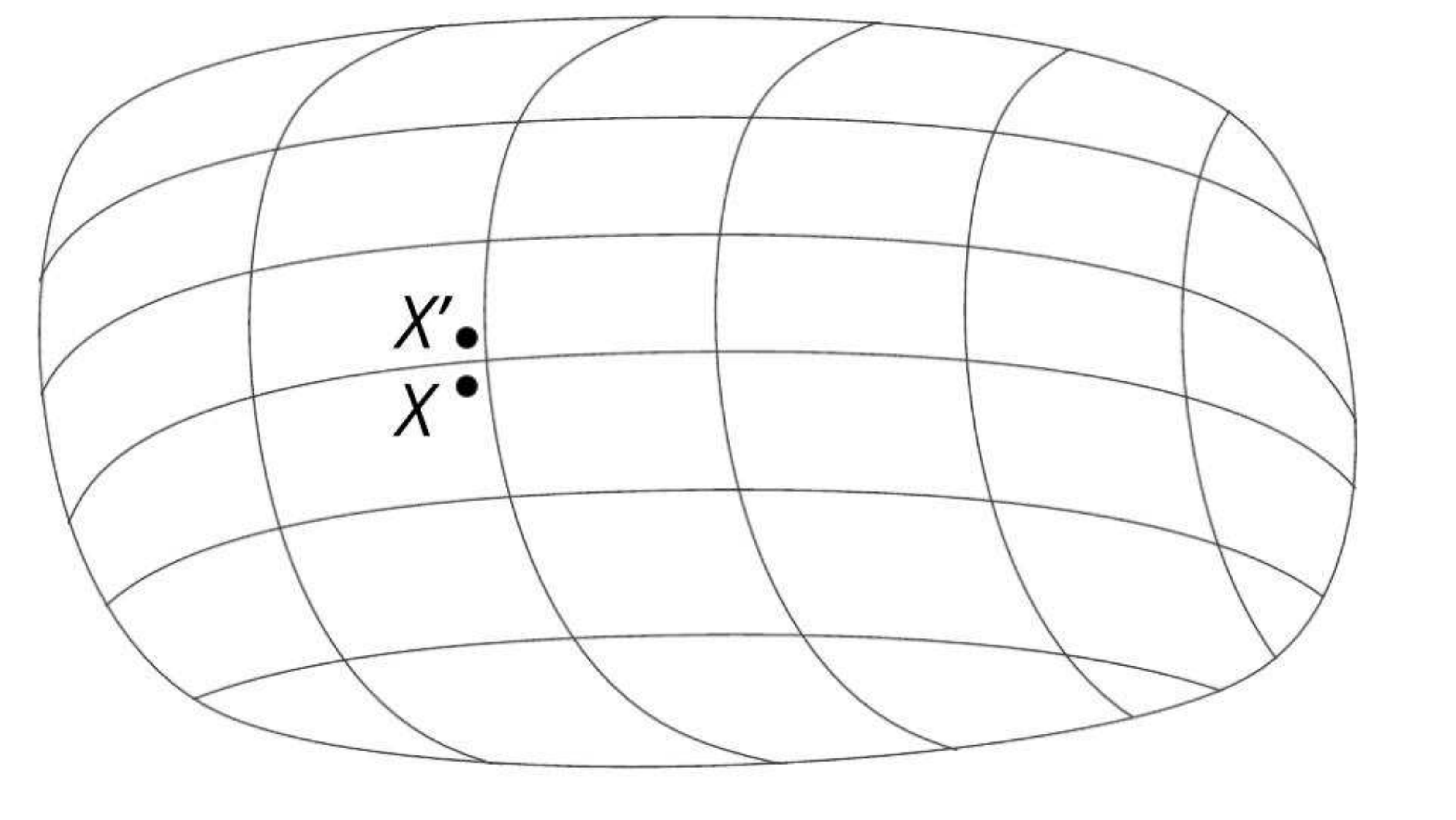}\\
\caption{The \textit{block-edge problem}. Objects $X$ and $X'$ in two catalogs. Even if corresponding to the same source, they can be discarded by the algorithm, since they belong to two different blocks of the sky partition.}\label{fig:block-edge}
\end{figure}

All the cross-matching algorithms based on a sky partitioning have to deal with the so-called block-edge problem, illustrated in Fig.~\ref{fig:block-edge}: the objects $X$ and $X'$ in different catalogs correspond to the same object but, falling in different pieces of the sky partition, the cross-matching algorithm is not able to identify the match. To solve this issue, it is necessary to add further steps to the pipeline, inevitably increasing the computational time. For example, the Zhao's tool, \citep{zhao2009}, expands a Healpix block with an opportunely dimensioned border; instead, the algorithm described by \cite{du2014}, combining Healpix and HTM virtual indexing function shapes, is able to reduce the block-edge problem, because the lost objects in a partition may be different from one to another.

\section{$C^3$ design and architecture}\label{sect:c3design}
$C^{3}$ is a command-line open-source Python script, designed and developed to perform a wide range of cross-matching types among astrophysical catalogs. The tool is able to be easily executed as a stand-alone process or integrated within any generic data reduction/analysis pipeline.
Based on a specialized sky partitioning function, its high-performance capability is ensured by making use of the multi-core parallel processing paradigm. It is designed to deal with massive catalogs in different formats, with the maximum flexibility given to the end-user, in terms of catalog parameters, file formats, coordinates and cross-matching functions.

In $C^{3}$ different functional cases and matching criteria have been implemented, as well as the most used join function types. It also works with the most common catalog formats, with or without header: Flexible Image Transport System (FITS, version tabular), American Standard Code for Information Interchange (ASCII, ordinary text, i.e. space separated values), Comma Separated Values (CSV), Virtual Observatory Table (VOTable, XML based) and with two kinds of coordinate system, equatorial and galactic, by using STILTS in combination with some standard Python libraries, namely \textit{NumPy}\footnote{\url{http://www.numpy.org/}} \citep{vanderwalt2011}, and \textit{PyFITS}\footnote{PyFITS is a product of the Space Telescope Science Institute, which is operated by AURA for NASA. \url{http://www.stsci.edu/institute/software\textunderscore hardware/pyfits}}.

\begin{figure*}
\centering
  \includegraphics[width=16.5cm]{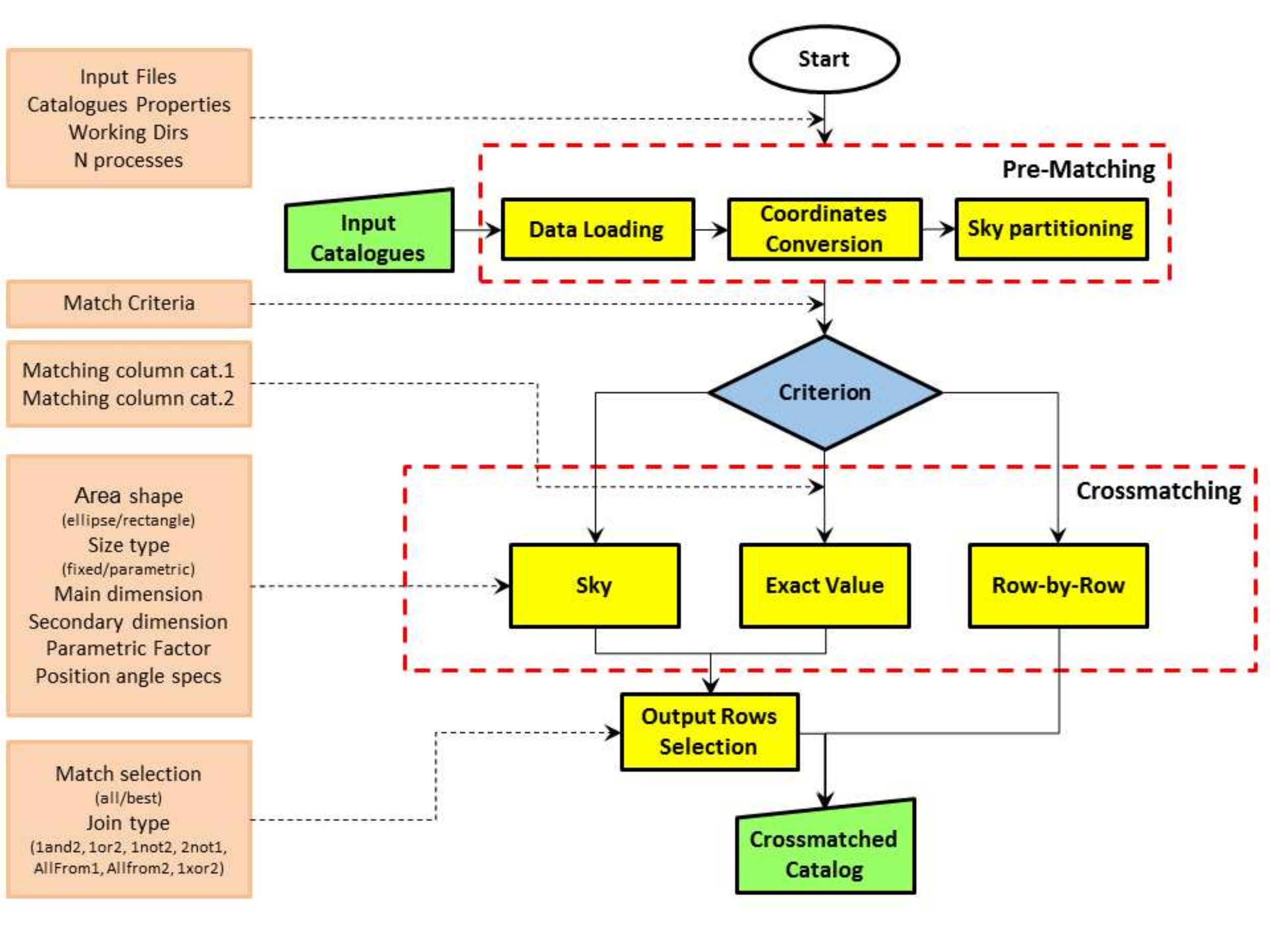}\\
\caption{Flowchart of the $C^{3}$ tool. The configuration requires few parameters (square panels on the left), according to the chosen match criterion. Currently three different functional cases are available (\textit{Sky}, \textit{Exact Value}, \textit{Row-by-Row}). The pipeline foresees a pre-matching step in order to prepare data for the multiprocess cross-matching phase.}\label{fig:flowchart}
\end{figure*}

Despite the general purpose of the tool, reflected in a variety of possible functional cases, $C^{3}$ is easy to use and to configure through few lines in a single configuration file. Main features of $C^{3}$ are the following:

\begin{enumerate}
 \item \textit{Command line}: $C^{3}$ is a command-line tool. It can be used as stand-alone process or integrated within more complex pipelines;
 \item \textit{Python compatibility}: compatible with Python 2.7.x and 3.4.x (up to the latest version currently available, $3.5$);
 \item \textit{Multi-platform}: $C^{3}$ has been tested on Ubuntu Linux $14.04$, Windows $7$ and $10$, Mac OS and Fedora;
 \item \textit{Multi-process}: the cross-matching process has been developed to run by using a multi-core parallel processing paradigm;
 \item \textit{User-friendliness}: the tool is very simple to configure and to use; it requires only a configuration file, described in Sec.~\ref{sect:config}.
\end{enumerate}

The internal cross-matching mechanism is based on the sky partitioning into cells, whose dimensions are determined by the parameters used to match the catalogs. The sky partitioning procedure is described in \ref{sect:preproc}. The Fig.~\ref{fig:flowchart} shows the most relevant features of the $C^{3}$ processing flow and the user parameters available at each stage.

\subsection{Functional cases}\label{sect:usecases}

As mentioned before, the user can run $C^{3}$ to match two input catalogs by choosing among three different functional cases:
\begin{enumerate}
 \item \textit{Sky}: the cross-match is done within sky areas (elliptical or rectangular) defined by the celestial coordinates taken from catalog parameters;
 \item \textit{Exact Value}: two objects are matched if they have the same value for a pair of columns (one for each catalog) defined by the user;
  \item \textit{Row-by-Row}: match done on a same row-ID of the two catalogs. The only requirement here is that the input catalogs must have the same number of records.
\end{enumerate}

The positional cross-match strategy of the $C^{3}$ method is based on the same concept of the Q-FULLTREE approach, an our tool introduced in \cite{becciani2015} and \cite{sciacca2016}: for each object of the first input catalog, it is possible to define an elliptical, circular or rectangular region centered on its coordinates, whose dimensions are limited by a fixed value or defined by specific catalog parameters. For instance, the two Full Width at Half Maximum (FWHM) values in the catalog can define the two semi-axes of an ellipse or the couple width and height of a rectangular region. It is also possible to have a circular region, by defining an elliptical area having equal dimensions. Once defined the region of interest, the next step is to search for sources of the second catalog within such region, by comparing their distance from the central object and the limits of the area (for instance, in the elliptical cross-match the limits are defined by the analytical equation of the ellipse).

\begin{figure}
\centering
  \includegraphics[width=8cm]{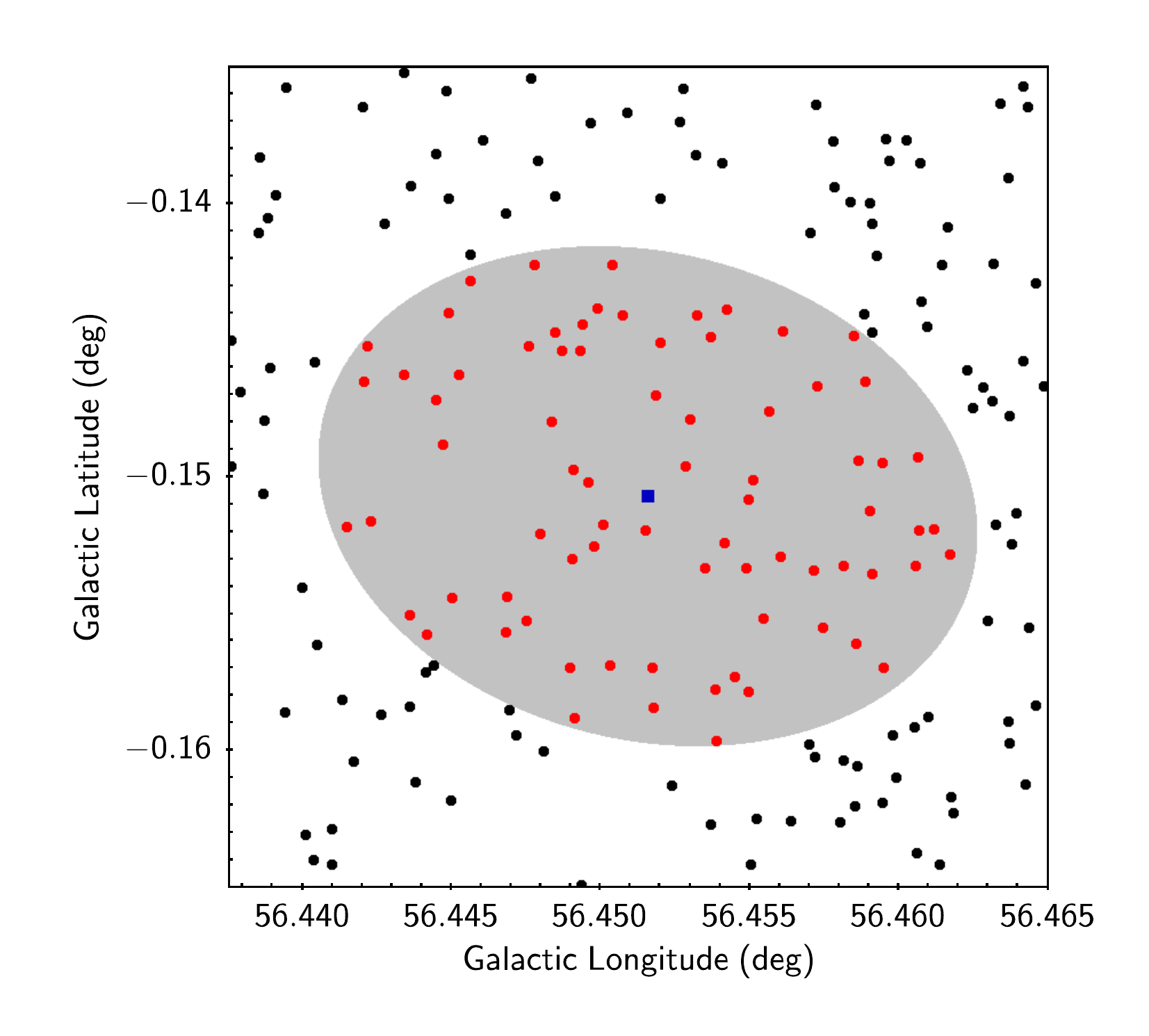}\\
\caption{Graphical representation of an elliptical cross-match between two catalogs: the gray ellipse represents the matching region defined by the FWHMs referred to an object of first catalog (squared dot in the center of the ellipse); all other points (belonging to the second catalog), that fall into the region defined by the ellipse (red or light gray dots), are matching with the central object.}\label{fig:crossmatch}
\end{figure}

In the \textit{Sky} functional case, the user can set additional parameters in order to characterize the matching region and the properties of the input catalogs. In particular, the user may define:

\begin{enumerate}
 \item the shape (elliptical or rectangular) of the matching area, i.e. the region, centered on one of the matching sources, in which to search the objects of the second catalog;
 \item the dimensions of the searching area. They can be defined by fixed values (in arcseconds) or by parametric values coming from the catalog. Moreover, the region can be rotated by a position angle (defined as fixed value or by a specific column present in the catalog);
 \item the coordinate system for each catalog (galactic, icrs, fk4, fk5) and its units (degrees, radians, sexagesimal), as well as the columns containing information about position and designation of the sources.
\end{enumerate}

An example of graphical representation of an elliptical cross-match is shown in Fig.~\ref{fig:crossmatch}.

In the \textit{Exact Value} case, the user has to define only which columns (one for each input catalog) have to be matched, while in the most simple \textit{Row-by-Row} case no particular configuration is needed.

\subsection{Match selection and join types}\label{sect:join}

\begin{figure}
\centering
\subfigure[1 and 2]{\includegraphics[width=2.3cm]{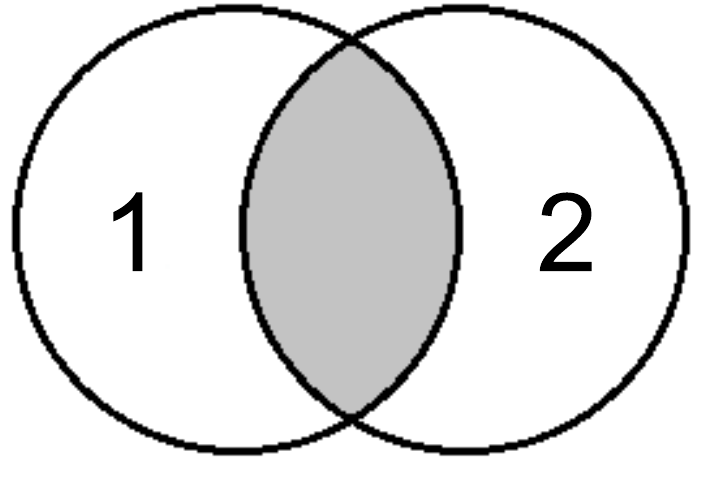}}
\hspace{2.5mm}
\subfigure[1 or 2]{\includegraphics[width=2.3cm]{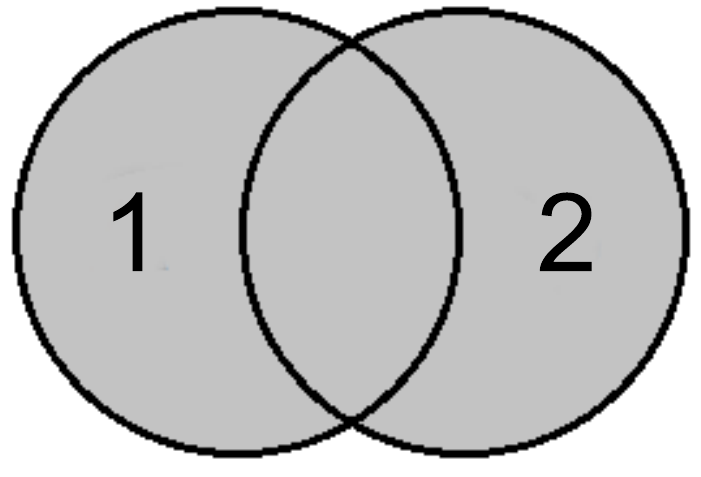}}\\
\subfigure[All from 1]{\includegraphics[width=2.3cm]{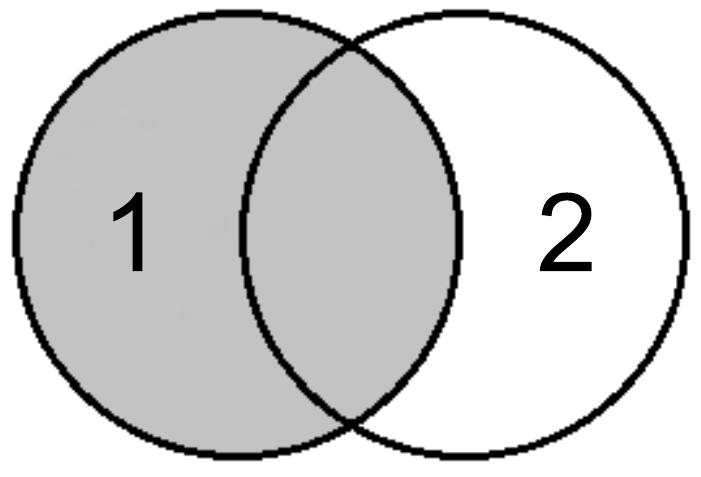}}
\hspace{2.5mm}
\subfigure[All from 2]{\includegraphics[width=2.3cm]{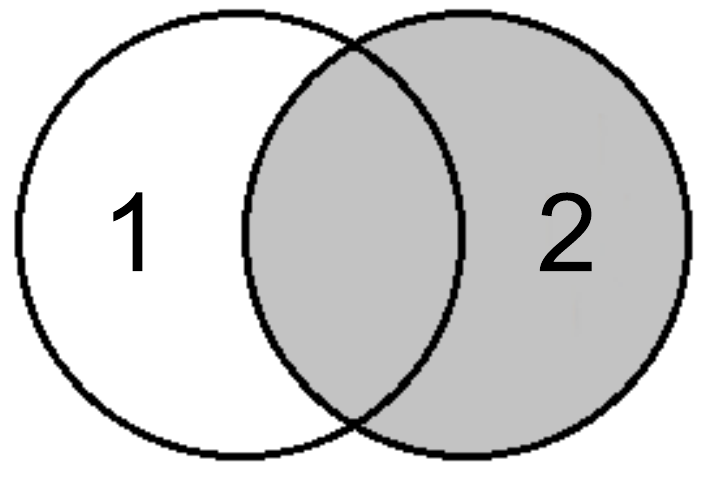}}
\hspace{2.5mm}
\subfigure[1 not 2]{\includegraphics[width=2.3cm]{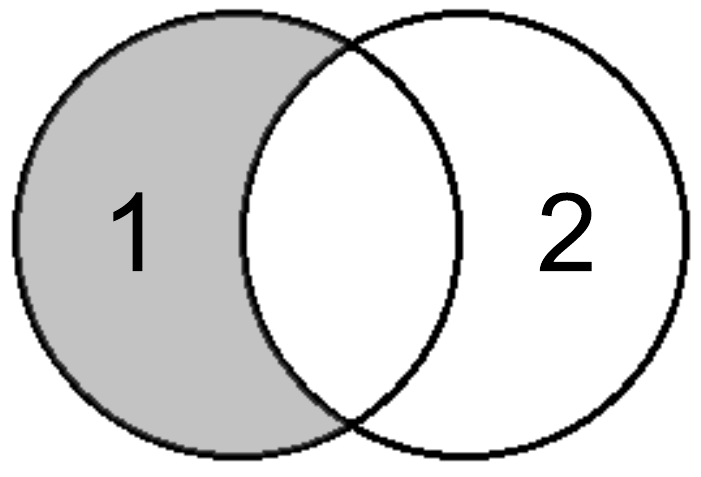}}\\
\subfigure[2 not 1]{\includegraphics[width=2.3cm]{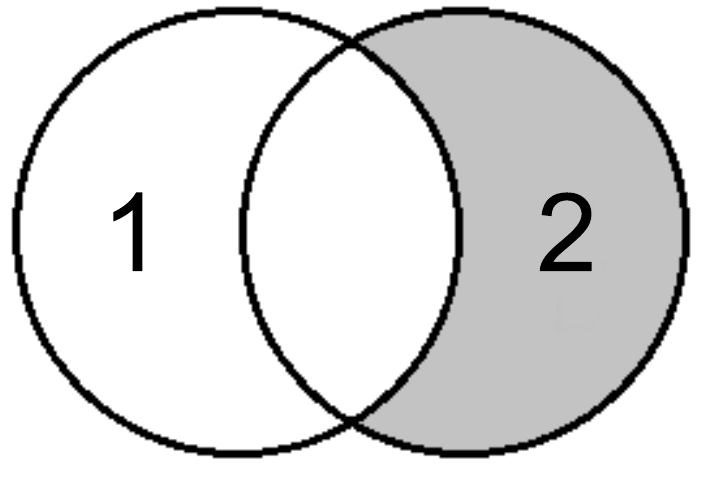}}
\hspace{2.5mm}
\subfigure[1 xor 2]{\includegraphics[width=2.3cm]{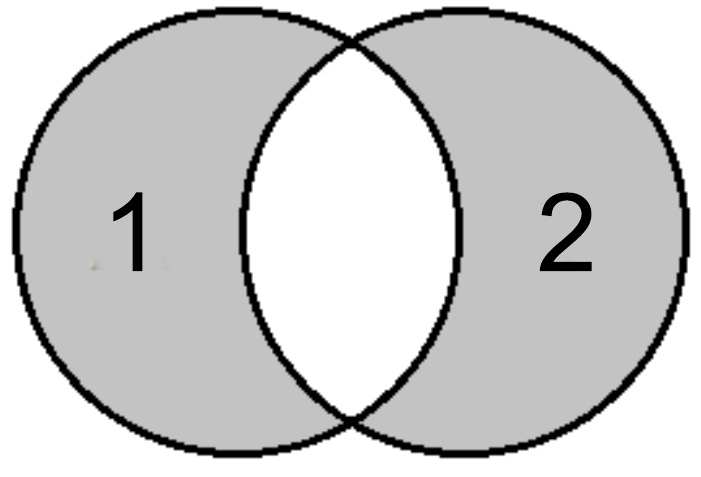}}

\caption{Join types available for $C^{3}$ output. Panel (a): rows both in catalog $1$ and catalog $2$ ($1$ and $2$); (b) all rows of catalog $1$ and catalog $2$ ($1$ or $2$); (c) all rows of catalog $1$ (all from $1$); (d) all rows of catalog $2$ (all from $2$); (e) rows in catalog $1$ not matched with catalog $2$ ($1$ not $2$); (f) rows in catalog $2$ not matched with catalog $1$ ($2$ not $1$); (g) rows from the catalog $1$ not having matches in the catalog $2$ and viceversa ($1$ xor $2$).}\label{fig:joins}
\end{figure}

$C^{3}$ produces a file containing the results of the cross-match, consisting into a series of rows, corresponding to the matching objects. In the case of \textit{Exact value} and \textit{Sky} options, the user can define the conditions to be satisfied by the matched rows to be stored in the output. First, it is possible to retrieve, for each source, all the matches or only the best pairs (in the sense of closest objects, according to the match selection criterion); then, the user can choose different join possibilities (in Fig.~\ref{fig:joins} the graphical representation of available joins is shown):

\begin{description}
 \item[$1$ and $2$] only rows having an entry in both input catalogs, (Fig.~\ref{fig:joins}a);
 \item[$1$ or $2$] all rows, matched and unmatched, from both input catalogs, (Fig.~\ref{fig:joins}b);
 \item[All from $1$ (All from $2$)] all matched rows from catalog $1$ (or $2$), together with the unmatched rows from catalog $1$ (or $2$), (Fig.~\ref{fig:joins}c-d);
 \item[$1$ not $2$ ($2$ not $1$)] all the rows of catalog $1$ (or $2$) without matches in the catalog $2$ (or $1$), (Fig.~\ref{fig:joins}e-f);
 \item[$1$ xor $2$] the ``exclusive or'' of the match - i.e. only rows from the catalog $1$ not having matches in the catalog $2$ and viceversa, (Fig.~\ref{fig:joins}g).
\end{description}

\subsection{Execution phases}
Any experiment with the $C^{3}$ tool is based on two main phases (see Fig.~\ref{fig:flowchart}):

\begin{enumerate}
 \item \textit{Pre-matching:} this is the first task performed by $C^{3}$ during execution. The tool manipulates input catalogs to extract the required information and prepare them to the further analysis;
 \item \textit{Matching:} after data preparation, $C^{3}$ performs the matching according to the criteria defined in the configuration file.
\end{enumerate}

Finally, the results are stored in a file, according to the match criterion described in Sec.~\ref{sect:join}, and all the temporary data are automatically deleted.

\subsubsection{Pre-matching}\label{sect:preproc}

This is the preliminary task performed by $C^{3}$ execution. During the pre-matching phase, $C^{3}$ performs a series of preparatory manipulations on input data. First of all, a validity check of the configuration parameters and input files. Then it is necessary to split the data sets in order to parallelize the matching phase and improve the performance.
In the \textit{Exact Value} functional case only the first input catalog will be split, while in the \textit{Sky} case both data sets will be partitioned in subsets. In the latter case, $C^{3}$ makes always use of galactic coordinates expressed in degrees, thus converting them accordingly if expressed in different format.

\begin{figure}
\centering
\subfigure[]{\includegraphics[width=3.8cm]{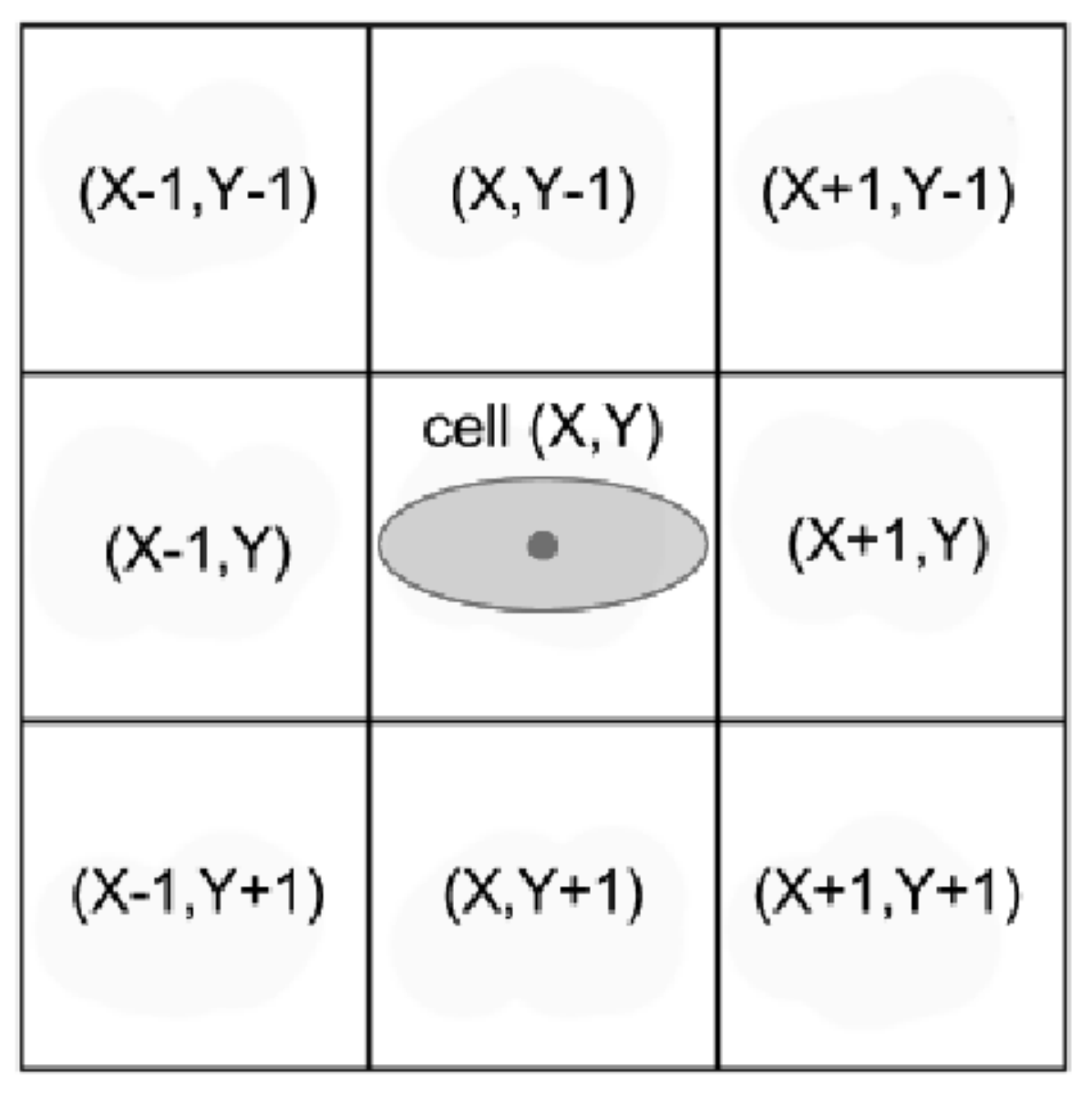}}
\hspace{5mm}
\subfigure[]{\includegraphics[width=3.8cm]{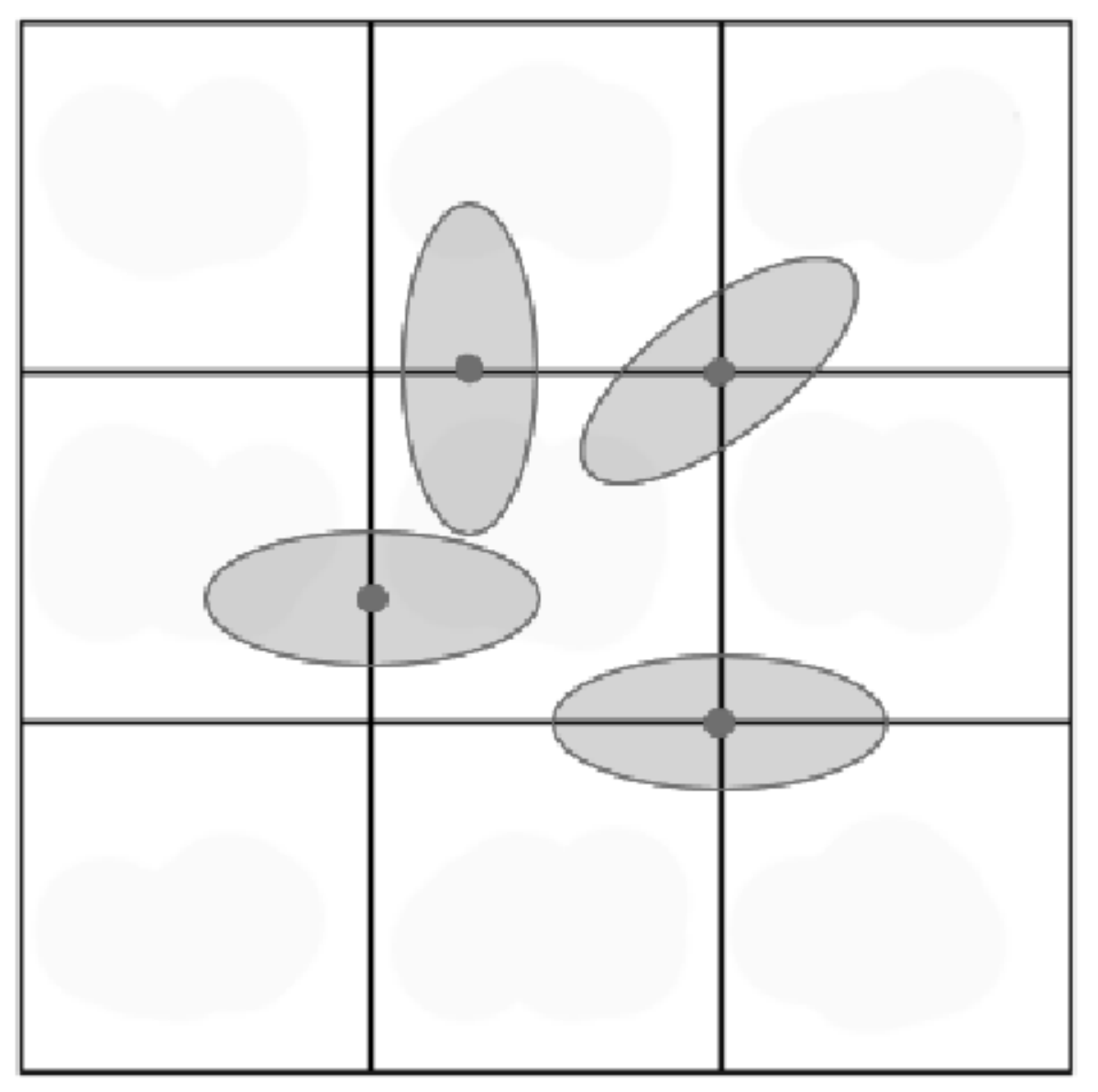}}

\caption{The $C^{3}$ sky partitioning method. The sky is partitioned in cells whose dimensions are determined by the maximum value assumed by the main dimension of the matching area or by the \textit{minimum partition cell size} parameter (panel a). Each object of the second catalog is assigned to a cell: a match between a source and the ellipse defined by the first catalog object can happen only in the nine cells surrounding it (panel b).}\label{fig:partitioning}
\end{figure}

When required, the two catalogs are split in the following way: in the first catalog all the entries are divided in groups, whose number depends on the multi-processing settings (see Sec.~\ref{sect:config}), since each process is assigned to one group; in the second catalog the sky region defined by the data set is divided into square cells, by assigning a cell to each entry, according to its coordinates (Fig.~\ref{fig:partitioning}).

We used the Python multiprocess module to overcome the GIL problem, by devoting particular care to the granularity of data to be handled in parallel. This implies that the concurrent processes do not need to share resources, since each process receives different files in input (group of object of the 1st catalog and cells) and produces its own output. Finally the results are merged to produce the final output.

The partitioning procedure on the second catalog is based on the dimensions of the matching areas: the size of the unit cell is defined by the maximum dimension that the elliptical matching regions can assume. If the ``Size type'' is ``parametric'', then the maximum value of the columns indicated in the configuration is used as cell size; in the case of ``fixed'' values, the size of the cell will be the maximum of the two values defined in the configuration (Fig.~\ref{fig:partitioning}a). In order to optimize the performance, the size of the unit cell cannot be less than a threshold value, namely the \textit{minimum partition cell size}, which the user has to set through the configuration file. The threshold on the cell size is required in order to avoid the risk to divide the sky in too many small areas (each one corresponding to a file stored on the disk), which could slow down the cross-matching phase performance. In Sec.~\ref{sect:optimization} we illustrated a method to optimize such parameter as well as the number of processes to use, according to the hosting machine properties.

Once the partitioning is defined, each object of the second catalog is assigned to one cell, according to its coordinates. Having defined the cells, the boundaries of an elliptical region associated to an object can fall at maximum in the eight cells surrounding the one including the object, as shown in Fig.~\ref{fig:partitioning}b. This prevents the block-edge problem previously introduced.

\subsubsection{Matching}\label{sect:matching}

Once the data have been properly re-arranged, the cross-match analysis can start. In the \textit{Row-by-Row} case, each row of the first catalog is simply merged with the corresponding row of the second data set through a serial procedure. In the other functional cases, the cross-matching procedure has been designed and implemented to run by using parallel processing, i.e. by assigning to each parallel process one group generated in the previous phase. In the \textit{Exact Value} case, each object of the group is compared with all the records of the second catalog and matched according to the conditions defined in the configuration file.

In the \textit{Sky} functional case, the matching procedure is slightly more complex. As described in Sec.~\ref{sect:usecases}, the cross-match at the basis of the $C^{3}$ method is based on the relative position of two objects: for each object of the first input catalog, $C^{3}$ defines the elliptical/rectangular region centered on its coordinates and dimensions. Therefore a source of the second catalog is matched if it falls within such region.

In practice, as explained in the pre-matching phase, having identified a specific cell for each object of a group, this information is used to define the minimum region around the object used for the matching analysis. The described choice to set the dimensions of the cells ensures that, if a source matches with the object, it must lie in the nine cells surrounding the object (also known as Moore's neighborhood, \citealt{gray2003}, see also Fig.~\ref{fig:partitioning}b). Therefore it is sufficient to cross-match an object of a group only with the sources falling in nine cells.

\begin{figure*}
\centering
\includegraphics[width=15cm]{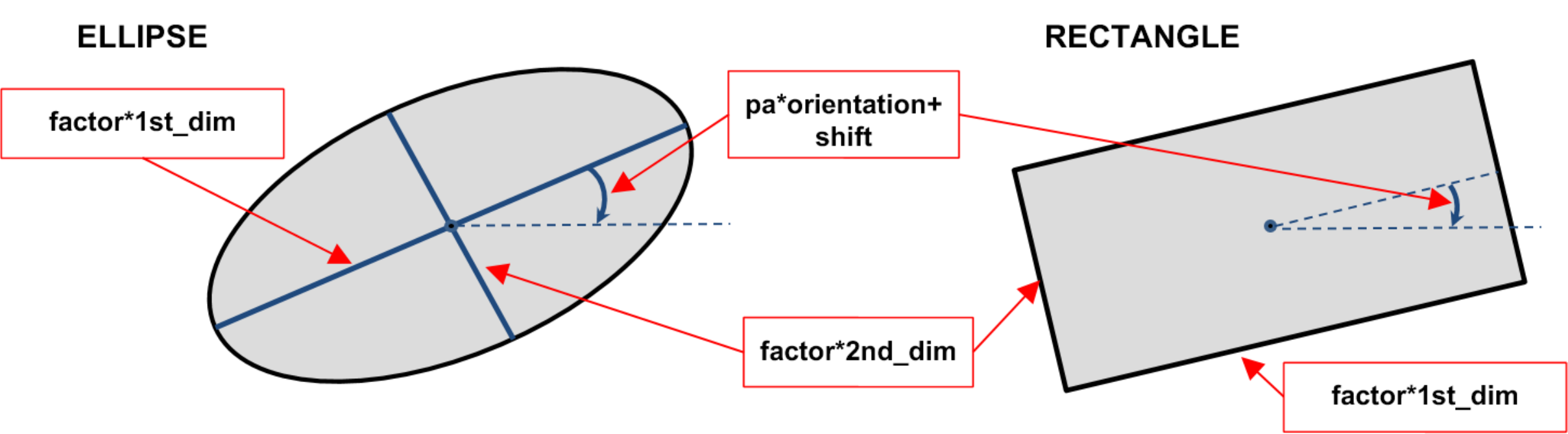}

\caption{Configuration of $C^{3}$ Matching Area: it can be elliptical (circular as special case) or rectangular; its dimensions, defined in the configuration file as \textit{matching area 1st and 2nd dimension}, represent the ellipse axes or width and height of the rectangle, multiplied, in the case of parametric size type, by a user defined \textit{parametric factor}; the position angle is characterized by a value (in degree) and two additional parameters, respectively, \textit{orientation} and \textit{shift}.}\label{fig:pa}
\end{figure*}

In the \textit{Sky} functional case, $C^{3}$ performs a cross-matching of objects lying within an elliptical, circular or rectangular area, centered on the sources of the first input catalog. The matching area is characterized by $6$ configuration parameters defining its shape, dimensions and orientation. In Fig.~\ref{fig:pa} is depicted a graphical representation of two matching areas (elliptical and rectangular) with the indication of its parameters.

In particular, to define the orientation of the matching area, $C^{3}$ requires two further parameters besides the offset and the value of the position angle, representing its orientation. The position angle, indeed, is referred, by default, to the greatest axis of the matching area with a clockwise orientation. The two additional parameters give the possibility to indicate, respectively, the correct orientation (clockwise/counterclockwise) and a shift angle (in degrees).

Finally, the results of the cross-matching are stored in a file, containing the concatenation of all the columns of the input catalogs referred to the matched rows. In the \textit{Sky} functional case the column reporting the separation distance between the two matching objects is also included.

\section{Configuration}\label{sect:config}

The tool $C^{3}$ is interfaced with the user through a single configuration file, to be properly edited just before the execution of any experiment. If the catalogs do not contain the source's Designation/ID information, $C^{3}$ will automatically assign an incremental row-ID to each entry as object designation.

For the \textit{Sky} functional case, assuming that both input catalogs contain the columns reporting the object coordinates, $C^{3}$ is able to work with galactic and equatorial (icrs, fk4, fk5) coordinate systems, expressed in the following units: degrees, radians or sexagesimal.

If the user wants to use catalog information to define the matching region (for instance, the FWHMs or a radius defined by the instrumental resolution), obviously the first input catalog must contain such data. The position angle value/column is, on the contrary, an optional information (default is 0$^o$, clockwise).

$C^{3}$ is conceived for a community as wide as possible, hence it has been designed in order to satisfy the requirement of user-friendliness.  Therefore, the configuration phase is limited to the editing of a setup file,\footnote{$C^{3}$ can also automatically generate a dummy configuration file that could be used as template.} containing all the information required to run $C^{3}$. This file is structured in sections, identified by square brackets: the first two are required, while the others depend on the particular use case. In particular, the user has to provide the following information:

\begin{enumerate}
 \item the input files and their format (FITS, ASCII, CSV or VOTable);
 \item the name and paths of the temporary, log and output files;
 \item the match criterion, corresponding to one of the functional cases (\textit{Sky, Exact Value, Row-by-Row}).
\end{enumerate}

$C^{3}$ gives also the possibility to set the number of processes running in parallel, through an optional parameter which has as default the number of cores of the working machine (minus one left available for system auxiliary tasks).

\subsection{\textit{Sky} functional case}

The configuration for the \textit{Sky} functional case foresees the setup of specific parameters of the configuration file: those required to define the shape and dimensions of the matching area, the properties of the input catalogs already mentioned in Sec.~\ref{sect:usecases},  coordinate system, units as well as the column indexes for source coordinates and designation. In addition, a parameter characterizing the sky partitioning has to be set (see Sec.~\ref{sect:preproc} for further information).

The parameters useful to characterize the matching area are the following:

\begin{description}
 \item [Area shape] it can be elliptical or rectangular (circular is a special elliptical case);
 \item [Size type] the valid entries are \textit{fixed} or \textit{parametric}. In the first case, a fixed value will be used to determine the matching area; in the second, the dimensions and inclination of the matching area will be calculated by using catalog parameters;
 \item [First and second dimensions of matching area] the axes of the ellipse or width and height of the rectangular area. In case of fixed ``Size type'', they are decimal values (in arcsec), otherwise, they represent the index (integer) or name (string) of the columns containing the information to be used;
 \item [Parametric factor] it is required and used only in the case of parametric ``Size type''. It is a decimal number factor to be multiplied by the values used as dimensions, in order to increase or decrease the matching region, as well as useful to convert their format;
 \item [Pa column/value] it is the position angle value (in the ``fixed'' case, expressed in degrees) or the name/ID of the column containing the position angle information (in the ``parametric'' case);
 \item [Pa settings] the position angle, which in $C^{3}$ is referred, by default, to the main axis of the matching area (greatest) with a clockwise orientation. The two parameters defined here give the possibility to indicate the correct orientation (clockwise/counterclockwise) and a shift angle (in degrees).
\end{description}

The user has also to specify which rows must be included in the output file, by setting the two parameters indicating the match selection and the join type, as described in Sec.~\ref{sect:join}.

\subsection{\textit{Exact Value} functional case}

For the \textit{Exact value} functional case it is required to set the name or id of the columns used for the match for both input files. The user has also to specify which rows must be included in the output file, by setting the two parameters indicating the match selection and the join type, as described in Sec.~\ref{sect:join}.

\subsection{\textit{Row-by-Row} functional case}

For the \textit{Row-by-Row} functional case, no other settings are required. The only constrain is that both catalogs must have the same number of entries.

\section{Computational optimization procedure}\label{sect:optimization}

As reflected from the description of $C^{3}$, the choice of the best values for its internal parameters (in particular the number of parallel processes and the minimum cell size, introduced in Sec.~\ref{sect:preproc}), is crucial to obtain the best computational efficiency. This section is dedicated to show the importance of this choice, directly depending on the features of the hosting machine. In the following tests we used a computer equipped with an Intel(R) Core(TM) $i5-4460$, with one $3.20GHz$, $4-core$ CPU, $32$ GB of RAM and hosting Ubuntu Linux $14.04$ as operative system (OS) on a standard Hard Disk Drive. We proceeded by performing two different kinds of tests:

\begin{enumerate}
  \item a series of tests with a fixed value for the minimum cell size ($100''$) and different values of the number of parallel processes;
  \item a second series by using the best value of number of parallel processes found at previous step and different values for the minimum cell size.
\end{enumerate}

The configuration parameters used in this set of tests are reported in Table \ref{test1:settings}. The input data sets are two identical catalogs (CSV format) consisting of $100,000$ objects extracted from the UKIDSS GPS public data \citep{lucas2008}, in the range of galactic coordinates $l\in[50,60]$, $b\in[-1,1]$. Each record is composed by $125$ columns. The choice to cross-match a catalog with itself represent the worst case in terms of cross-matching computational time, since each object matches at least with itself.

By setting ``Match Selection'' as best and ``Join Type'' as 1 and 2 (see Table~\ref{test1:settings}), we obtained an output of $100,000$ objects matched with themselves as expected. We also performed all the tests by using a ``random shuffled`` version of the same input catalog, obtaining the same results. This demonstrates that the $C^{3}$ output is not affected by the particular order of data in the catalogs.

\begin{table}
\centering
\begin{tabular}{|c|c|}
\hline
\textbf{Parameter}	& \textbf{Value}\\   \hline
Area Shape 	        & Ellipse\\
Size Type          & Fixed\\
1st dimension (Major axis)          & $5''$\\
2nd dimension (Secondary axis)        & $5''$\\
Position Angle settings             & $0^o$\\
Coordinate System	            & Galactic (deg)\\
Match Selection            & best\\
Join type	        & 1 and 2\\
\hline
\end{tabular}
\caption{$C^{3}$ settings in the first set of tests performed to evaluate the impact of the number of parallel processes and the minimum cell size configuration parameters on the execution time. The choice of same dimensions for the ellipse axes was due to perform a fair comparison with STILTS and CDS-Xmatch, which allow only circular cross-matching.}
\label{test1:settings}
\end{table}

As expected, the number of parallel processes affects the partitioning of the first catalog. In particular, if a too large value is selected, it induces a negative impact on the computational efficiency, causing a bottleneck due to the higher frequency of disk access.

The results of these tests, shown in Table~\ref{test1:results}, confirm that the best choice of the concurrent processes is not the highest one. In fact, although the serial case ($N=1$) is obviously the worst result, the computational time reaches the minimum with $N=256$, from which it starts to increase. The overall \textit{speedup} achieved in the best case is $\sim7\times$ with respect to the serial case.

The computational time of the pre-matching phase appears almost constant in all tests, because this portion of the $C^{3}$ code is not parallel in the current version of the tool. The small time fluctuations of such phase are due to the unpredictable status of the hosting machine during the tests. The output creation phase, depending on the number of temporary files produced by each concurrent cross-matching process, reaches an almost constant value, mainly imposed by the serial nature of this phase.

\begin{table*}
\scriptsize
\setlength{\tabcolsep}{6pt} 
\centering
\begin{tabular}{|c|c|c|c|c|c|}
\hline
\textbf{TestID}	&\textbf{N processes}	&\textbf{Pre-matching}	&\textbf{Cross-matching}	&\textbf{Output creation}	&\textbf{Total }\\
&	&\textbf{time (s)}	&\textbf{time (s)}	&\textbf{time (s)}	&\textbf{time (s)}\\	\hline
  NP1 & 1 & 29 & 11 & 412 & 452\\
  NP4 & 4 & 28 & 3 & 108 & 139\\
  NP8 & 8 & 28 & 3 & 72 & 102\\
  NP16 & 16 & 28 & 3 & 54 & 85\\
  NP20 & 20 & 28 & 3 & 50 & 81\\
  NP32 & 32 & 28 & 3 & 45 & 76\\
  NP64 & 64 & 28 & 3 & 40 & 71\\
  NP100 & 100 & 28 & 3 & 40 & 71\\
  NP128 & 128 & 28 & 3 & 39 & 70\\
  NP256 & 256 & 28 & 4 & 37 & 69\\
  NP512 & 512 & 28 & 4 & 38 & 70\\
  NP1024 & 1024 & 28 & 5 & 38 & 72\\
  NP2048 & 2048 & 28 & 8 & 39 & 74\\
  NP2560 & 2560 & 28 & 11 & 38 & 77\\
  NP3072 & 3072 & 28 & 13 & 38 & 79\\ \hline
\end{tabular}
\caption{The computational time of the whole process (column $6$) and of each phase of the tool execution (columns from $3$ to $5$), for experiments with the same configuration but different number of parallel processes (column $2$). Here the minimum cell size is fixed to $100$ arcsecs.
The input data sets are two identical catalogs consisting of $100,000$ objects extracted from the UKIDSS GPS public data. Each record is composed by $125$ columns.}
\label{test1:results}
\end{table*}

Once the best number of concurrent processes has been chosen, we proceeded by looking for the value of the minimum cell size that provides the best result in terms of computational time. The number of subsets, in which the first input catalog has to be divided, depends on the number of parallel processes, while the minimum cell size determines the granularity of the second catalog, corresponding to the resolution of the sky partitioning. A too high cell size implies a partition with few large areas; a too small value causes the generation of a too large number of regions with very few objects.

The parameters used in this set of tests are the same of the previous step (see Table~\ref{test1:settings}), with the number of concurrent processes fixed to $N=256$. We decided to vary the cell size between $25''$ and $200''$. The results of the test are reported in Table~\ref{test2:results}.

\begin{table*}
\scriptsize
\setlength{\tabcolsep}{6pt} 
\centering
\begin{tabular}{|c|c|c|c|c|c|}
\hline
\textbf{TestID}	&\textbf{Cell size}	&\textbf{Pre-matching}	&\textbf{Cross-matching}	&\textbf{Output creation}	&\textbf{Total}\\

	&\textbf{(arcseconds)}	&\textbf{time (s)}	&\textbf{time (s)}	&\textbf{time (s)}	&\textbf{time (s)}\\	\hline

  TH25 & 25 & 38 & 4 & 36 & 77\\
  TH50 & 50 & 31 & 3 & 41 & 76\\
  TH75 & 75 & 29 & 3 & 41 & 74\\
  TH100 & 100 & 28 & 3 & 40 & 71\\
  TH125 & 125 & 27 & 4 & 41 & 72\\
  TH150 & 150 & 27 & 5 & 42 & 73\\
  TH175 & 175 & 27 & 5 & 41 & 74\\
  TH200 & 200 & 26 & 6 & 41 & 74\\

\hline
\end{tabular}
\caption{The computational time of the whole process (column $6$) and of each phase of the tool execution (columns from $3$ to $5$), for experiments with the same configuration but different minimum cell size (column $2$). These tests have been done by fixing the number of parallel processes to $N=256$.
The input data sets are two identical catalogs consisting of $100,000$ objects extracted from the UKIDSS GPS public data. Each record is composed by $125$ columns.}
\label{test2:results}
\end{table*}

In this case, the pre-matching phase is, as expected, slightly affected by the choice of the cell size, because the region has to be divided in a different number of cells. While the computational time of the output phase, on the contrary, is not affected by the cell size. The duration of the cross-matching phase reaches a minimum at $50''$, $75''$ and $100''$, where the minimum of the total computational time, and hence the best performance, is reached using $100''$ as minimum cell size.

In more general terms, the described example demonstrates that, in order to obtain the best computational performance, the configuration requires a series of heuristics to reach the best compromise between the granularity of the parallel processing and the scheduling management of the OS. As rule of thumb, the best results can be obtained by choosing the number of parallel processes limited between $10$ and $100$ times the number of cores of the hosting machine.

For what concerns the minimum cell size, in the previous example we considered $20$ square degrees, with $100,000$ objects, thus a density of $\sim1$ object in $2600$ square arcsecs. Since the best results have been obtained with a cell size of $100''$, we obtained $\sim7$ objects per cell.
By extrapolating from our tests, the best cell size, conditioning the sky partitioning resolution, should be chosen between $2$ and $10$ sources per cell. Of course, this heuristic range depends on the specific density of the involved fields.

\section{Testing on astrophysical data}\label{sect:performances}

In order to validate the results of $C^{3}$ and evaluate its performance in terms of cross-matching reliability and computational time efficiency, we performed several tests on real data. In particular, two kinds of tests have been executed, both using the most complex functional case \textit{Sky}. The first set of tests (Sec.~\ref{sect:validation}), has been performed to validate the $C^{3}$ results in terms of matching capability, through a comparison with other available cross-matching tools, for instance, STILTS and CDS-Xmatch.
The second set of tests (Sec.~\ref{sect:comparison}), has been used to evaluate the computational time efficiency, by varying the dimensions of the input data sets (both in terms of rows and columns), again through a comparison with the other tools.

\subsection{Cross-matching validation tests}\label{sect:validation}

In order to assess the reliability of the cross-matches produced by $C^{3}$, we performed an intensive test campaign. In this section we report the most significative examples which well represent the behavior of the tool. This set of tests has been performed by applying our tool on two data sets with variable number of objects and by comparing the results with those obtained by other applications representative of different paradigms: stand-alone command-line (STILTS, release 3.0-7), GUI (TOPCAT, release 4.2.3) and web application (CDS-XMatch\footnote{\url{http://cdsxmatch.u-strasbg.fr/xmatch}}).

\begin{table*}[!ht] \centering \scriptsize
\setlength{\tabcolsep}{6pt} 
\begin{tabular}{|c|c|c|c|c|}
\hline
  \textbf{ID} &
  \textbf{N$_{input1}$} &
  \textbf{N$_{input2}$} &
  \textbf{$C^{3}$, CDS-XMatch,} &
  \textbf{$C^{3}$, STILTS/TOPCAT }\\
  & & & \textbf{STILTS/TOPCAT (all)} & \textbf{(best)}
  \\ \hline

  T1 & 1000 & 1000 & 0 & 0\\
  T2 & 1000 & 10,000 & 1 & 1\\
  T3 & 1000 & 100,000 & 5 & 5\\
  T4 & 1000 & 1,000,000 & 116 & 116\\
  T5 & 10,000 & 1000 & 0 & 0\\
  T6 & 10,000 & 10,000 & 14 & 14\\
  T7 & 10,000 & 100,000 & 116 & 116\\
  T8 & 10,000 & 1,000,000 & 1260 & 1248\\
  T9 & 100,000 & 1000 & 12 & 12\\
  T10 & 100,000 & 10,000 & 136 & 136\\
  T11 & 100,000 & 100,000 & 1212 & 1211\\
  T12 & 100,000 & 1,000,000 & 12,711 & 12,758\\
  T13 & 1,000,000 & 1000 & 141 & 137\\
  T14 & 1,000,000 & 10,000 & 1295 & 1267\\
  T15 & 1,000,000 & 100,000 & 12,701 & 12,416\\
  T16 & 1,000,000 & 1,000,000 & 126,965 & 123,261\\
  T17 & 10,000,000 & 1000 & 191 & 169\\
  T18 & 10,000,000 & 10,000 & 1995 & 1755\\
  T19 & 10,000,000 & 100,000 & 19,717 & 17,235\\
  T20 & 10,000,000 & 1,000,000 & 196,310 & 171,775\\

\hline
\end{tabular}
\caption{Cross-matching results of $C^{3}$, STILTS/TOPCAT and CDS-Xmatch, for different dimensions of input catalogs (columns $2$ and $3$). Column $4$ reports the number of matches of the three tools in the case of \textit{all} matching selection criterion (tests T17-T20 have not been performed for CDS-Xmatch), while column $5$ reports the matches found using the \textit{best} criterion. In both cases all tools provided exactly the same number of matches in the whole set of tests.}
\label{tab:matchres}
\end{table*}

The first input catalog has been extracted by the UKIDSS GPS data in the range of galactic coordinates $l\in[40,50]$, $b\in[-1,1]$, while the second input catalog has been extracted by the GLIMPSE \textit{Spitzer} Data, (\citealt{benjamin2003} and \citealt{churchwell2009}), in the same range of coordinates. From each catalog, different subsets with variable number of objects have been extracted. In particular, data sets with, respectively, $1000$, $10,000$, $100,000$, $1,000,000$ and $10,000,000$ objects have been created from the first catalog, while, from second catalog, data sets with $1000$, $10,000$, $100,000$ and $1,000,000$ rows have been extracted. Then, each subset of first catalog has been cross-matched with all the subsets of the second catalog. For uniformity of comparison, due to the limitations imposed by CDS-XMatch in terms of available disk space, it has been necessary to limit to only $3$ the number of columns for all the subsets involved in the tests performed to compare C$^3$ and CDS-XMatch (for instance, ID and galactic coordinates). For the same reason, the data set with $10^7$ rows has not been used in the comparison between C$^3$ and CDS-XMatch.

The common internal configuration used in these tests is shown in Table~\ref{test1:settings}, except for the ``Match Selection`` parameter. There was, in fact, the necessity to set it to \textit{all} for uniformity of comparison with the CDS-Xmatch tool (which makes available only this option). Then the \textit{best} type has been used to compare $C^{3}$ with STILTS and TOPCAT. Furthermore, in all the tests, the number of parallel processes was set to $256$ and the minimum cell size to $100''$, corresponding to the best conditions found in the optimization process of $C^{3}$ (see Sec.~\ref{sect:optimization}). Finally, we chose same dimensions of the ellipse axes in order to be aligned with other tools, which allow only circular cross-matching areas.

Concerning the comparison among $C^{3}$ and the three mentioned tools, in the cases of both \textit{all} and \textit{best} types of matching selection, all tools provided exactly the same number of matches in the whole set of tests, thus confirming the reliability of $C^{3}$ with respect to other tools (Table~\ref{tab:matchres}).\footnote{For uniformity of comparison, due to the limitations imposed by CDS-XMatch, the data set with $10^7$ rows has not been used.}

\subsection{Performance tests}\label{sect:comparison}

In terms of computational efficiency, $C^{3}$ has been evaluated by comparing the computational time of its cross-matching phase with the other tools. The pre-matching and output creation steps have been excluded from the comparison, because strongly dependent on the host computing infrastructure.
The other configuration parameters have been left unchanged (Table~\ref{test1:settings}). The complete setup for the described experiments is reported in the Appendix.

In Fig.~\ref{fig:c3vsSTrows} we show the computational time of the cross-matching phase for $C^{3}$ and STILTS, as function of the incremental number of rows (objects) in the first catalog, and by varying the size of the second catalog in four cases, spanning from $1000$ to $1,000,000$ rows. In all diagrams, it appears evident the difference between the two tools, becoming particularly relevant with increasing amounts of data.


\begin{figure*}
\centering
\subfigure[]{\includegraphics[width=8.6cm]{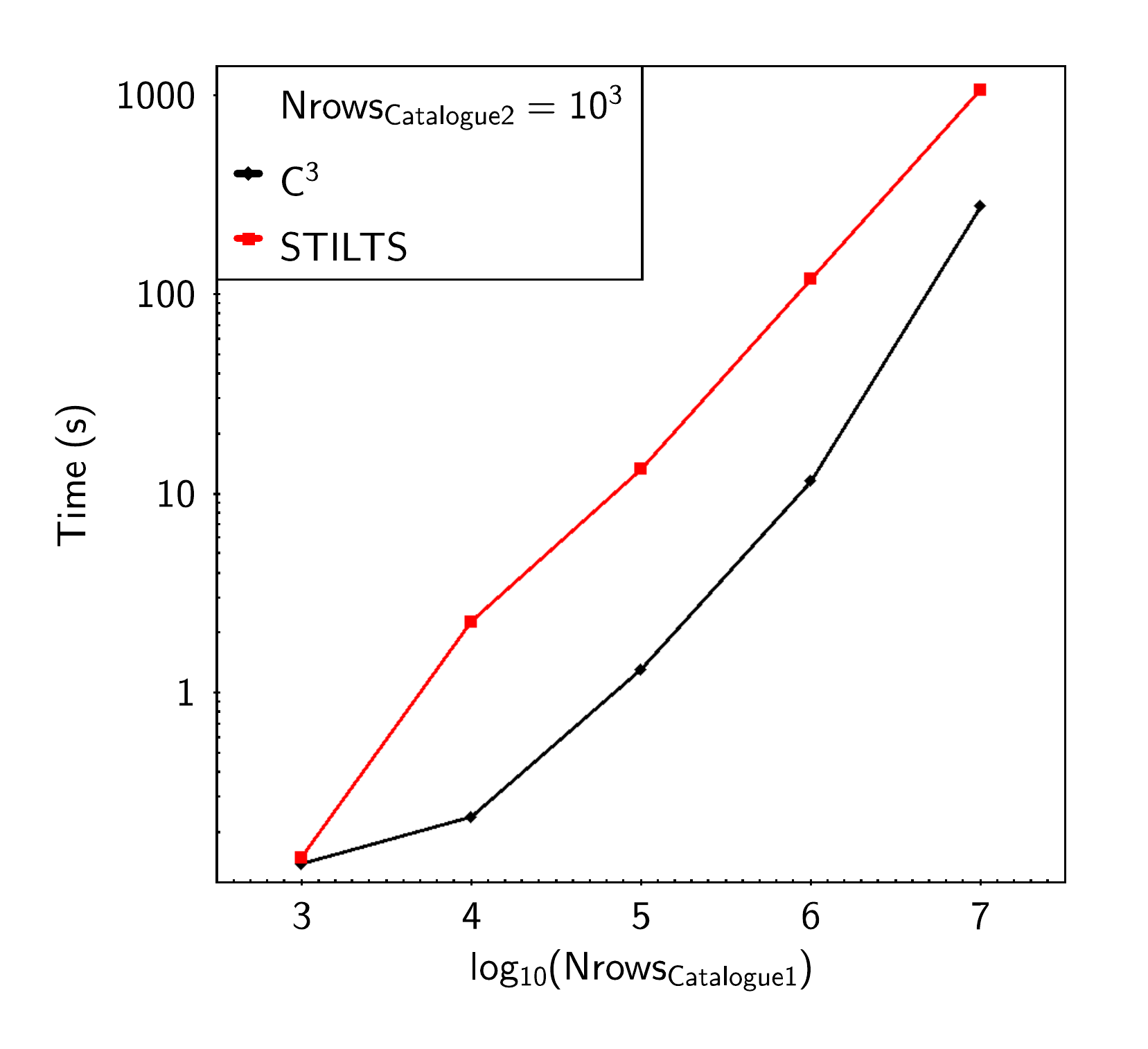}}
\hspace{5mm}
\subfigure[]{\includegraphics[width=8.6cm]{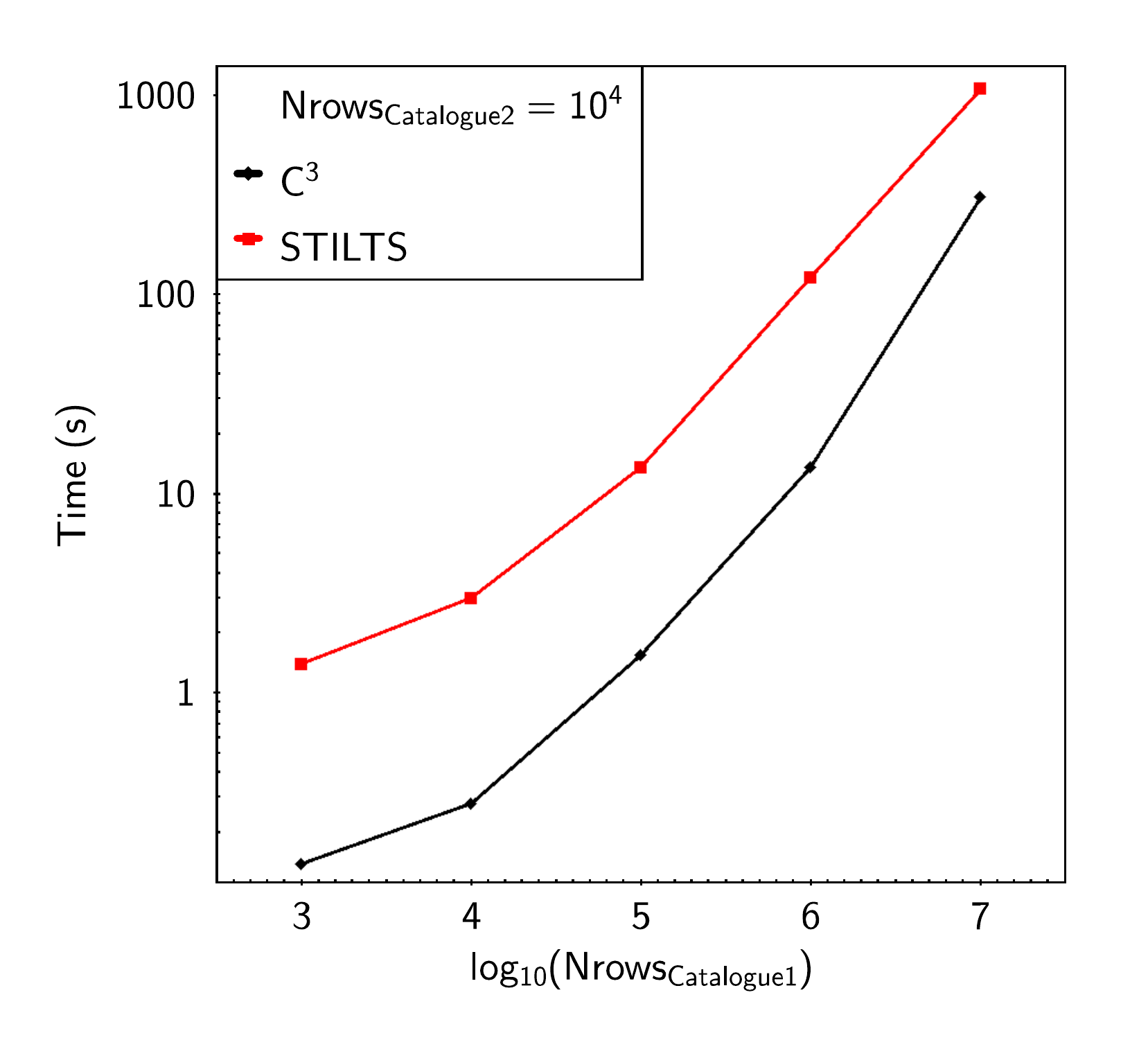}}
\hspace{5mm}
\subfigure[]{\includegraphics[width=8.6cm]{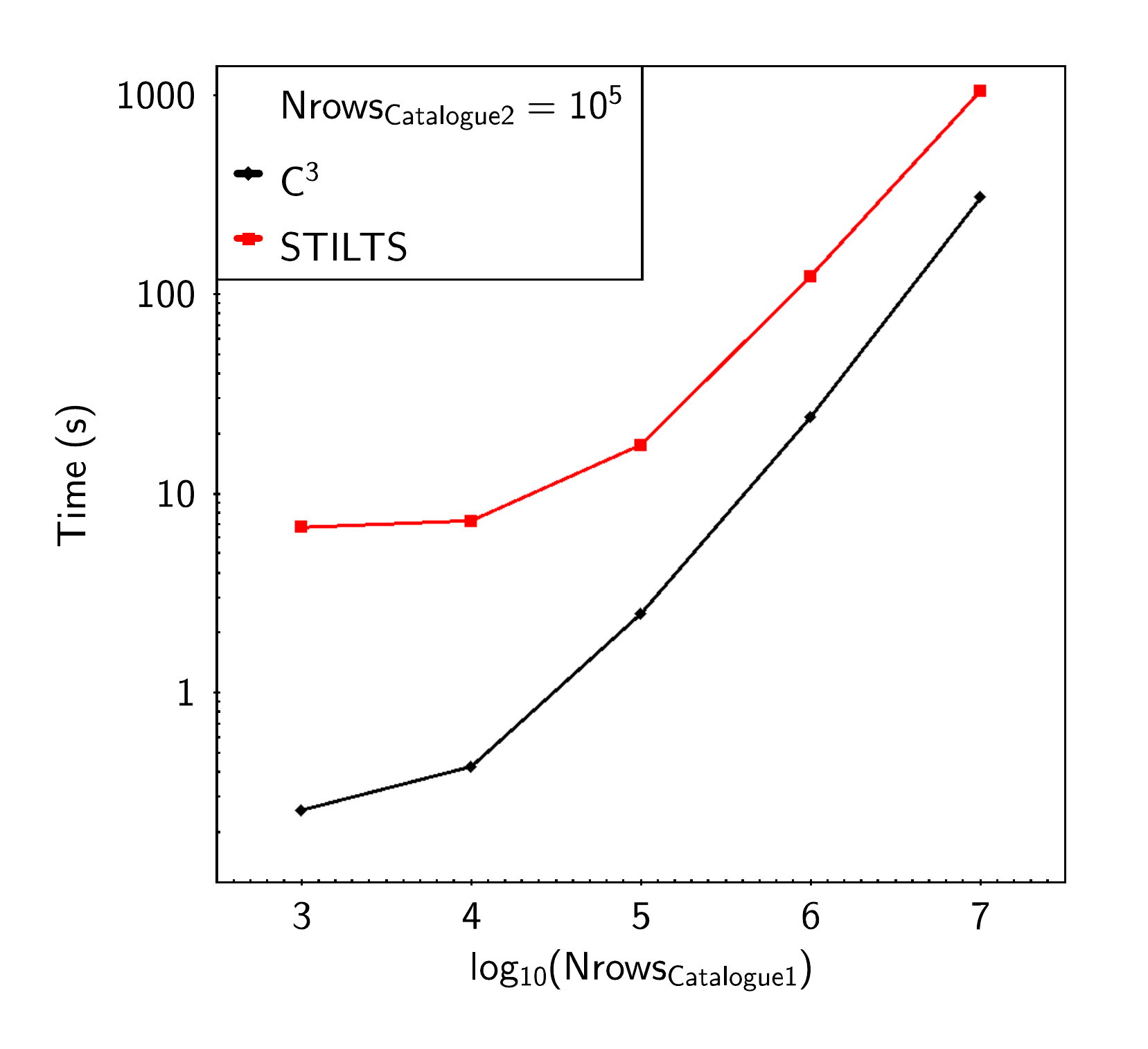}}
\hspace{5mm}
\subfigure[]{\includegraphics[width=8.6cm]{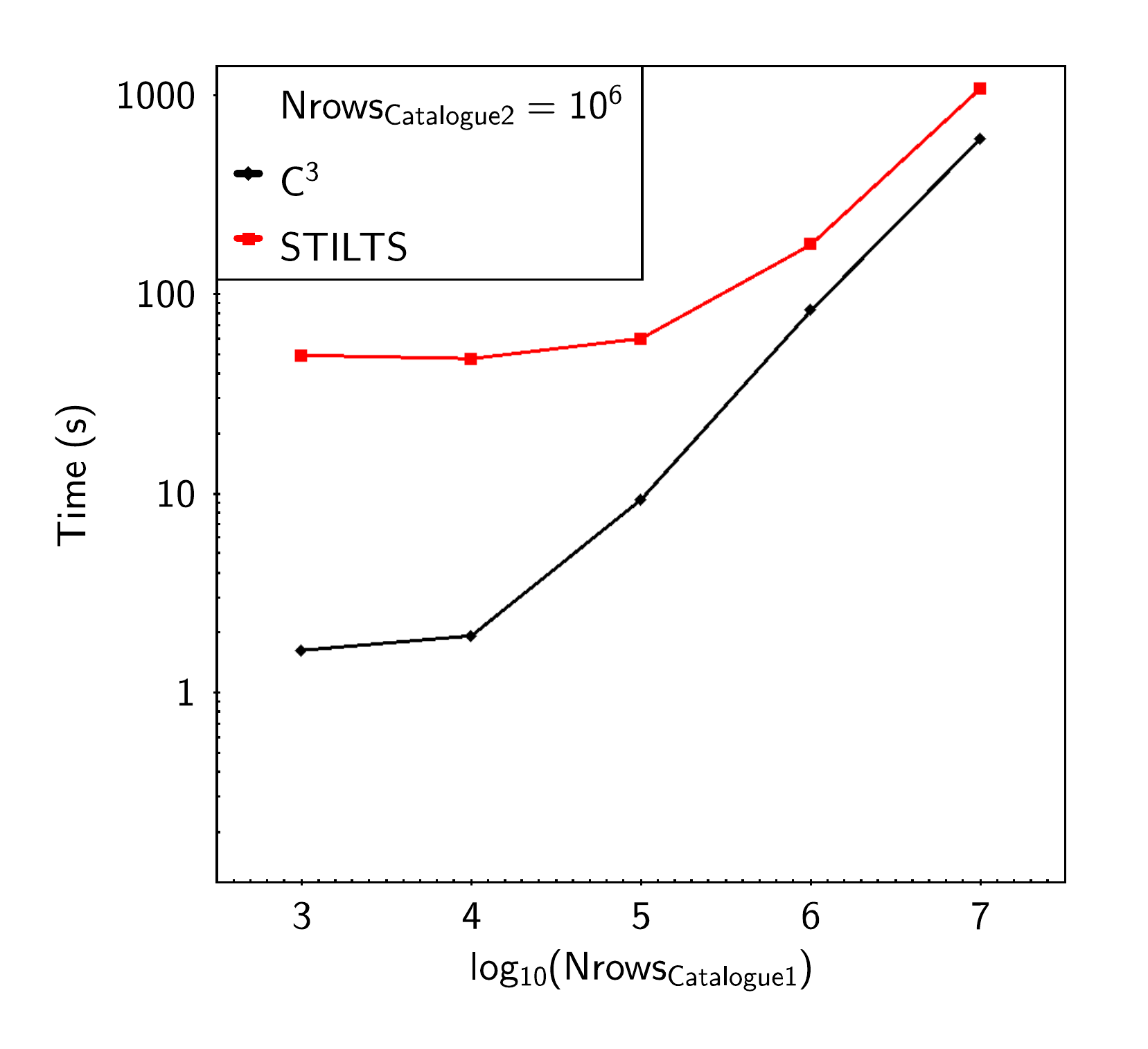}}

\caption{Computational time trends as function of the number of rows of the first input catalog for the $C^{3}$ cross-matching phase (black) and STILTS (red or gray) for four different dimensions of the second catalog: (a) $1000$ rows, (b) $10,000$ rows, (c) $100,000$ rows, (d) $1,000,000$ rows.}\label{fig:c3vsSTrows}
\end{figure*}

\begin{figure*}
\centering
\includegraphics[width=15cm]{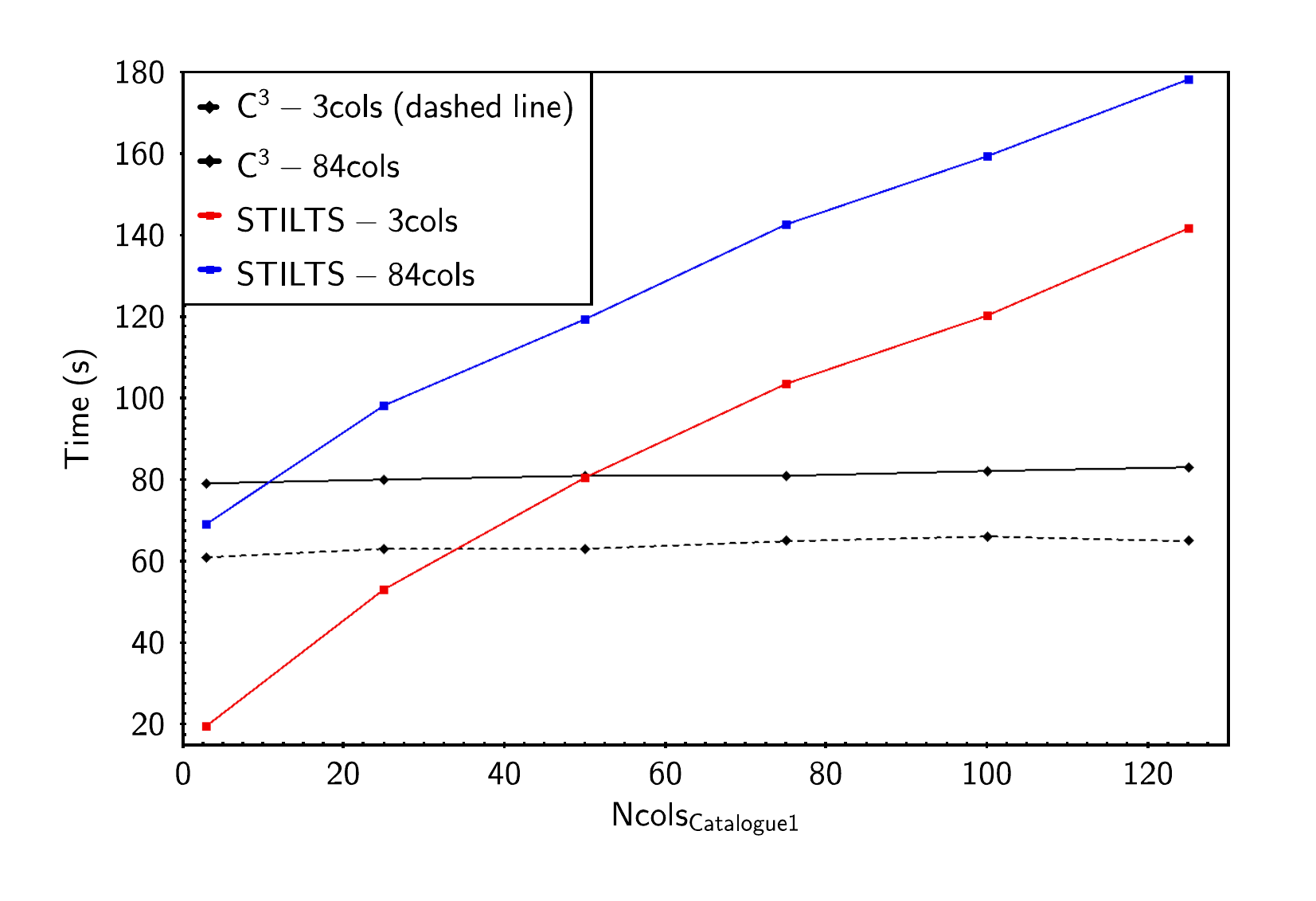}

\caption{Computational time as function of the number of columns of the first input catalog for the $C^{3}$ cross-matching phase and STILTS, considering a second catalog with $3$ (black dashed line for $C^3$, red or light gray line for STILTS) and $84$ (black line for $C^3$ and blue or dark gray for STILTS) columns.}\label{fig:c3vsSTcols}
\end{figure*}

In the second set of tests performed on the $C^{3}$ cross-matching phase and STILTS, the computational time has been evaluated as function of the incremental number of columns of the first catalog (from the minimum required $3$ up to $125$, the maximum number of columns of catalog 1), and by fixing the number of columns of the second catalog in five cases, respectively, $3$, $20$, $40$, $60$ and $84$, which is the maximum number of columns for catalog 2. In terms of number of rows, in all cases both catalogs were fixed to $1,000,000$ of entries.
In Fig.~\ref{fig:c3vsSTcols} the results only for $3$ and $84$ columns of catalog 2 are reported, showing that $C^{3}$ is almost invariant to the increasing of columns, becoming indeed faster than STILTS from a certain amount of columns. Such trend is confirmed in all the other tests with different number of columns of the second catalog. This behavior appears particularly suitable in the case of massive catalogs. Finally, in the case of two FITS input files instead of CSV files, STILTS computational time as function of the number of columns is constant and slightly faster than $C^3$.

\begin{figure*}
\centering
\subfigure[]{\includegraphics[width=8.6cm]{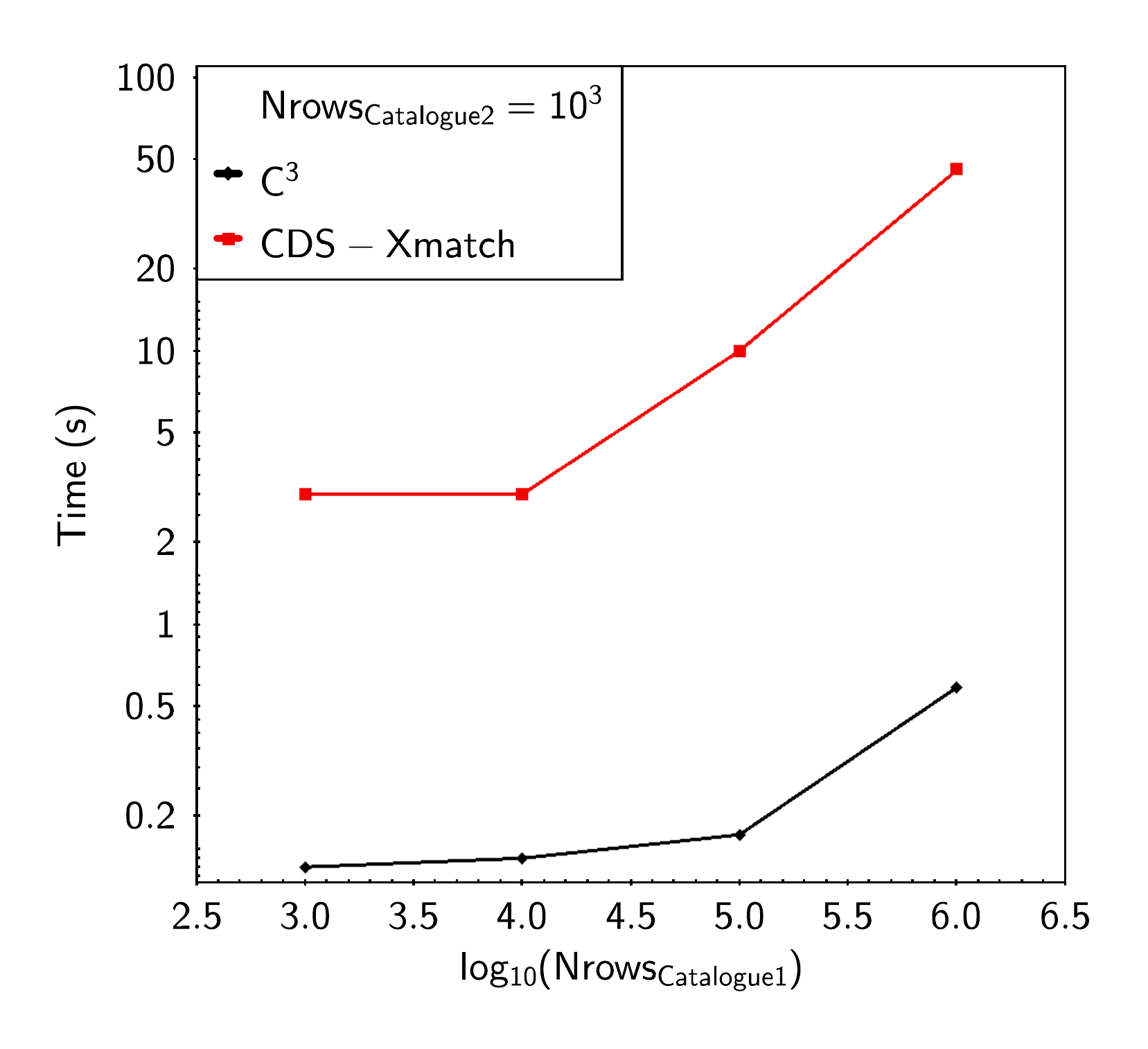}}
\hspace{5mm}
\subfigure[]{\includegraphics[width=8.6cm]{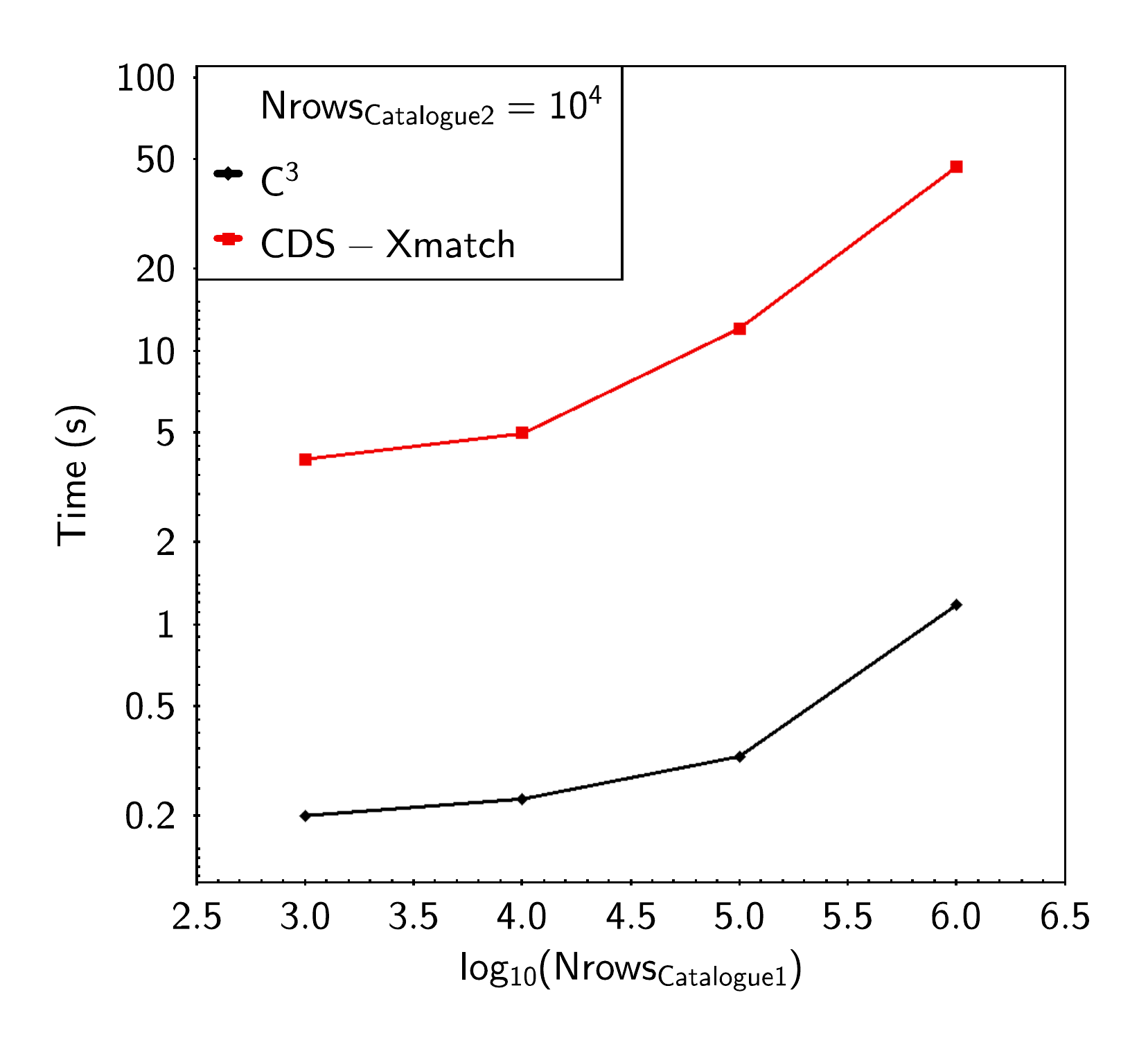}}
\hspace{5mm}
\subfigure[]{\includegraphics[width=8.6cm]{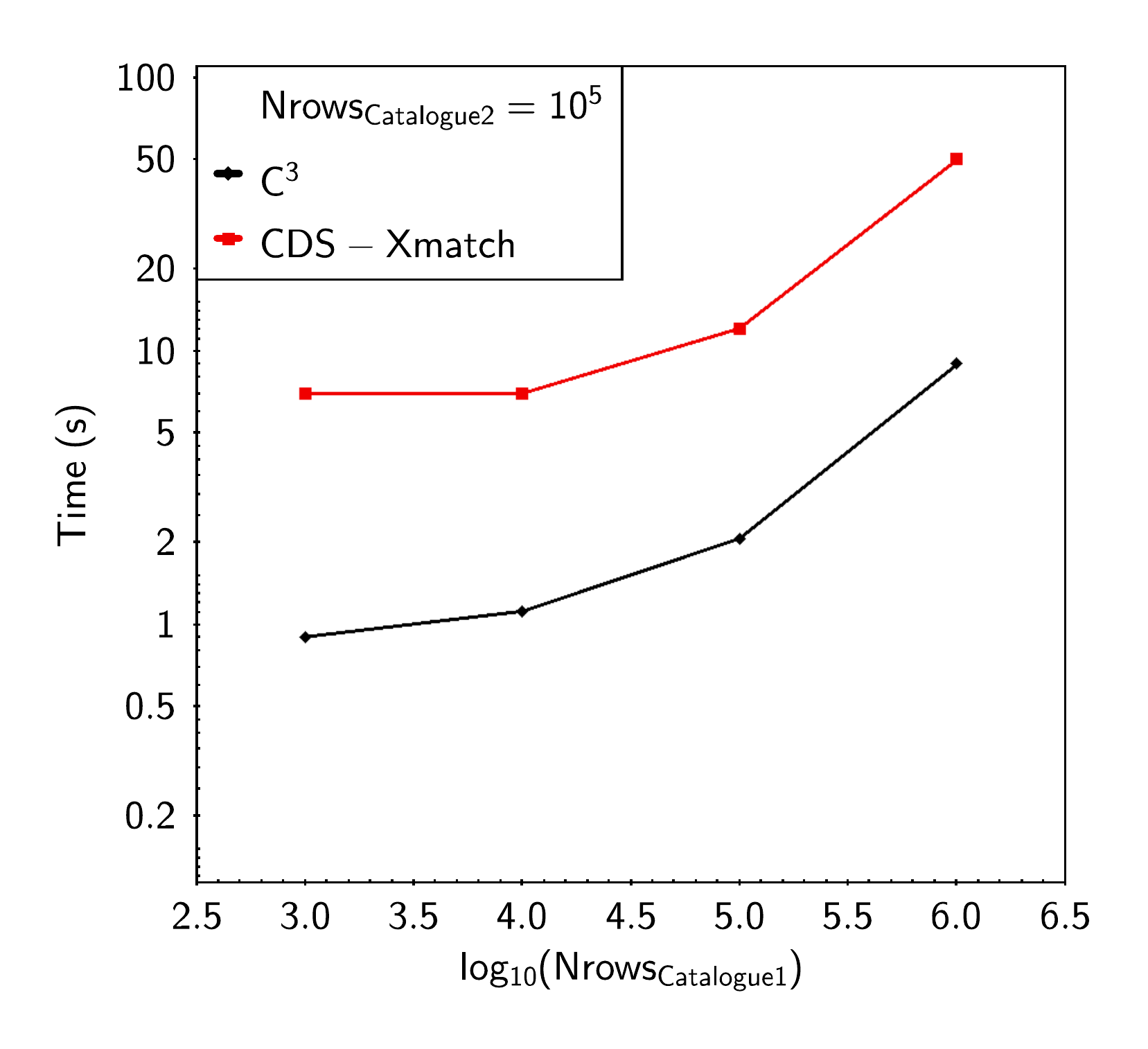}}
\hspace{5mm}
\subfigure[]{\includegraphics[width=8.6cm]{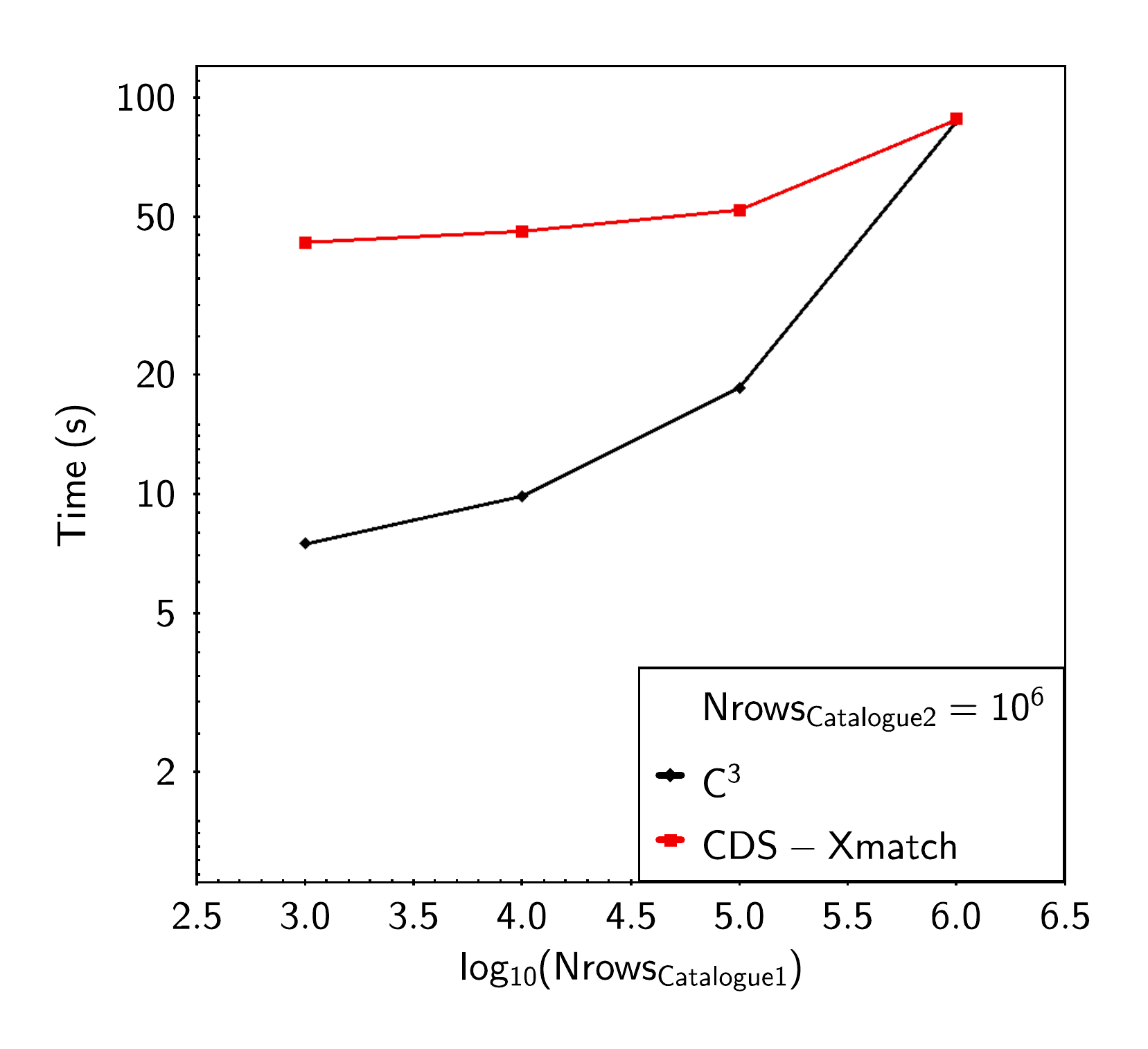}}

\caption{Computational time as function of the number of rows of the first input catalog for the $C^{3}$ cross-matching phase (black) and CDS-Xmatch (red or gray) for four different dimensions of the second catalog: (a) $1000$ rows, (b) $10,000$ rows, (c) $100,000$ rows, (d) $1,000,000$ rows. Considering only the cross-matching step of CDS-Xmatch, its performance are better than $C^3$.}
\label{fig:c3vsXMrows}
\end{figure*}

In the last series of tests, we compared the computational efficiency between the $C^{3}$ cross-matching phase and CDS-Xmatch. In this case, due to the limitation of the catalog size imposed by CDS-Xmatch, the tests have been performed by varying only the number of rows from $1000$ to $1,000,000$ as in the analogous tests with STILTS (except the test with $10,000,000$ rows), fixing the number of columns to $3$. Moreover, in this case, the cross-matching phase of $C^{3}$ has been compared with the duration of the phase \textit{execution} of the CDS-Xmatch experiment, thus ignoring latency time due to the job submission, strongly depending on the network status and the state of the job queue, but taking into account the whole job execution. The results, reported in Fig.~\ref{fig:c3vsXMrows}, show a better performance of $C^{3}$, although less evident when both catalogs are highly increasing their dimensions, where the differences due to the different hardware features become more relevant.

At the end of the test campaign, two other kinds of tests have been performed: (i) the verification of the portability of $C^{3}$ on different OSs and (ii) an analysis of the impact of different disk technology on the computing time efficiency of the tool.

In the first case, we noted, as expected, a decreasing of $C^{3}$ overall time performance on the Windows versions ($7$ and $10$), with respect to same tests executed on Linux versions (Ubuntu and Fedora) and MAC OS. On average $C^{3}$ execution was $\sim20$ times more efficient on Linux and MAC OS than Windows. This is most probably due to the different strategy of disk handling among various OSs, particularly critical for applications, like cross-matching tools, which make an intensive use of disk accesses.

This analysis induced us to compare two disk technologies: HDD (Hard Disk Drive) vs SSD (Solid State Disk). Both kinds of disks have been used on a sample of the tests previously described, revealing on average a not negligible increasing of computing time performance in the SSD case of $\sim1.4$ times with respect to HDD. For clarity, all test results presented in the previous sections have been performed on the same HDD.

\section{Conclusions and future developments}\label{sect:conclusion}
In this paper we have introduced $C^{3}$, a new scalable tool to cross-match astronomical data sets. It is a multi-platform command-line Python script, designed to provide the maximum flexibility to the end users in terms of choice about catalog properties (I/O formats and coordinates systems), shape and size of matching area and cross-matching type. Nevertheless, it is easy to configure, by compiling a single configuration file, and to execute as a stand-alone process or integrated within any generic data reduction/analysis pipeline.

In order to ensure the high-performance capability, the tool design has been based on the multi-core parallel processing paradigm and on a basic sky partitioning function to reduce the number of matches to check, thus decreasing the global computational time. Moreover, in order to reach the best performance, the user can tune on the specific needs the shape and orientation of the matching region, as well as tailor the tool configuration to the features of the hosting machine, by properly setting the number of concurrent processes and the resolution of sky partitioning. Although elliptical cross-match and the parametric handling of angular orientation and offset are known concepts in the astrophysical context, their availability in the presented command-line tool makes $C^{3}$ competitive in the context of public astronomical tools.

A test campaign, done on real public data, has been performed to scientifically validate the $C^{3}$ tool, showing a perfect agreement with other publicly available tools. The computing time efficiency has been also measured by comparing our tool with other applications, representative of different paradigms, from stand-alone command-line (STILTS) and graphical user interface (TOPCAT) to web applications (CDS-Xmatch). Such tests revealed the full comparable performance, in particular when input catalogs increase their size and dimensions.

For the next release of the tool, the work will be mainly focused on the optimization of the pre-matching and output creation phases, by applying the parallel processing paradigm in a more intensive way. Moreover, we are evaluating the possibility to improve the sky partitioning efficiency by optimizing the calculation of the minimum cell size, suitable also to avoid the block-edge problem.

The $C^{3}$ tool, \citep{riccio2016}, and the user guide are available at the page \url{http://dame.dsf.unina.it/c3.html}.

\section*{Acknowledgments}
The authors would like to thank the anonymous referee for extremely valuable comments and suggestions.
MB and SC acknowledge financial contribution from the agreement ASI/INAF I/023/12/1.
MB, AM and GR acknowledge financial contribution from the 7th European Framework Programme for Research Grant FP7-SPACE-2013-1, \textit{ViaLactea - The Milky Way as a Star Formation Engine}. MB and AM acknowledge the PRIN-INAF 2014 \textit{Glittering kaleidoscopes in the sky: the multifaceted nature and role of Galaxy Clusters}.


\appendix
\section{Configuration file example}\label{append}

This appendix reports the configuration file as used in the example described in Sec.~\ref{sect:comparison}.
The text preceded by the semicolon is a comment.

\begin{Verbatim}[commandchars=\\\{\}]
\textcolor{red}{[I/O Files]}
Input catalog 1: \textcolor{olive}{./input/ukidss.csv}
Format catalog 1: \textcolor{olive}{csv} \textcolor{blue}{;csv, fits, votable or ascii}
Input catalog 2: \textcolor{olive}{./input/glimpse.csv}
Format catalog 2: \textcolor{olive}{csv} \textcolor{blue}{;csv, fits, votable or ascii}
Output: \textcolor{olive}{./output/out.csv}
Output format: \textcolor{olive}{csv} \textcolor{blue}{;csv, fits, votable or ascii}
Log file: \textcolor{olive}{./output/out.log}
Stilts directory: \textcolor{olive}{./libs}
working directory: \textcolor{olive}{./tmp} \textcolor{blue}{;temporary directory, removed when completed}

\textcolor{red}{[Match Criteria]}
algorithm: \textcolor{olive}{sky} \textcolor{blue}{;sky, exact value, row-by-row}

\textcolor{red}{[Sky parameters]}
area shape: \textcolor{olive}{ellipse} \textcolor{blue}{;ellipse or rectangle}
size type: \textcolor{olive}{fixed} \textcolor{blue}{;parametric or fixed}
matching area first dimension: \textcolor{olive}{5} \textcolor{blue}{;arcsec for fixed type - column name/number for parametric type}
matching area second dimension: \textcolor{olive}{5} \textcolor{blue}{;arcsec for fixed type - column name/number for parametric type}
parametric factor: \textcolor{olive}{1} \textcolor{blue}{;multiplicative factor for dimension columns - required for parametric type}
pa column/value: \textcolor{olive}{0} \textcolor{blue}{;degrees for fixed type - column name/number for parametric type}
pa settings: \textcolor{olive}{clock, 0} \textcolor{blue}{;orientation (clock, counter), shift (degrees) -empty or default = clock,0}
Catalog 2 minimum partition cell size: \textcolor{olive}{100} \textcolor{blue}{;arcsec}

\textcolor{red}{[Catalog 1 Properties]}
coordinate system: \textcolor{olive}{galactic} \textcolor{blue}{;galactic, icrs, fk4, fk5}
coordinate units: \textcolor{olive}{deg} \textcolor{blue}{;degrees (or deg), radians (or rad), sexagesimal (or sex)}
glon/ra column: \textcolor{olive}{L} \textcolor{blue}{;column number or name - required for sky algorithm}
glat/dec column: \textcolor{olive}{B} \textcolor{blue}{;column number or name - required for sky algorithm}
designation column: \textcolor{olive}{SOURCEID} \textcolor{blue}{;column number or name - -1 for none}

\textcolor{red}{[Catalog 2 Properties]}
coordinate system: \textcolor{olive}{galactic} \textcolor{blue}{;galactic, icrs, fk4, fk5}
coordinate units: \textcolor{olive}{deg} \textcolor{blue}{;degrees (or deg), radians (or rad), sexagesimal (or sex)}
glon/ra column: \textcolor{olive}{l} \textcolor{blue}{;column number or name - required for sky algorithm}
glat/dec column: \textcolor{olive}{b} \textcolor{blue}{;column number or name - required for sky algorithm}
designation column: \textcolor{olive}{designation} \textcolor{blue}{;column number or name, -1 for none}

\textcolor{red}{[Threads Properties]}
thread limit: \textcolor{olive}{256} \textcolor{blue}{;maximum number of simultaneous threads (it depends on your machine)}

\textcolor{red}{[Output Rows]}
Match selection: \textcolor{olive}{all} \textcolor{blue}{;all or best}
Join type: \textcolor{olive}{1 and 2} \textcolor{blue}{;1 and 2, 1 or 2, all from 1, all from 2, 1 not 2, 2 not 1, 1 xor 2}
\end{Verbatim}
\clearpage
\end{document}